\documentclass[pre,floatfix,twocolumn,showpacs]{revtex4}
\usepackage{amsmath}
\usepackage{amsfonts}
\usepackage{mathrsfs}
\usepackage{epsfig}
\usepackage{graphicx}
\usepackage{dcolumn}
\usepackage{amsmath}
\usepackage{url}
\newcommand{\dd}[0]{{\textrm d}}

\begin{document}

\title{
Large variability in dynamical transitions in biological systems with quenched
disorder }
%Transition to sustained activity in a model of coupled cells:
%disorder-induced fluctuations, system-size dependence and scaling}
\author{
Jinshan Xu$^{1,2}$, Rajeev Singh$^3$, Nicolas Garnier$^1$, Sitabhra Sinha$^3$ and Alain Pumir$^1$}
\affiliation{$^1$Laboratoire de Physique, ENS-Lyon, CEDEX 69007, Lyon, France \\
$^2$ Department of Physics, East China Normal University, Shanghai, 20062, China \\
$^3$ The Institute of Mathematical Sciences, CIT Campus, Taramani,
Chennai 600113, India}
\date{\today}
%\ead{alain.pumir@ens-lyon.fr}
\begin{abstract}
Coherent oscillatory activity can 
arise spontaneously as a result of increased coupling 
in a system of excitable and passive 
cells, each being quiescent in isolation.
This can potentially
explain the appearance of spontaneous rhythmic contractions in the pregnant
uterus close to term.
%The model studied here involves excitable cells coupled on a $2$-dimensional 
%lattice, each of them coupled to a fluctuating number $n_p$ of passive cells, 
We investigate the transition to periodic activity using a model system 
comprising a lattice of excitable cells, each being coupled to a
variable number of passive cells, whose distribution defines  
a quenched
realization (replica) of spatial disorder. 
%We have studied the role of disorder
%in the transition to oscillatory behavior as the coupling between  
%adjacent excitable cells, and between passive and excitable 
%cells is varied, with the average 
%number of passive cells coupled to an excitable 
%cell for a given replica being kept fixed.
Close to the transition between quiescent state and sustained
oscillations in the system, 
we observe large fluctuations between different replicas induced by 
variations in the local density of passive cells around an excitable cell.
%We analyze this transition and its system-size dependence. 
We demonstrate that the disorder-induced fluctuations 
can be described in terms of a simple scaling 
relation which involves the strength of coupling between excitable cells,
the mean passive cell density, as well as the logarithm of the system size. 
%the difference of the average passive cell 
%density from the critical value at which the transition occurs.
Our results can be interpreted as suggesting that larger organs will
have greater variability in the onset of persistent activity.
%In particular, they may be more likely to exhibit oscillations 
%even prior to the transition point as a result of spatial fluctuations, 
%potentially implying that animals having bigger uteri will be at 
%higher risk of having significantly premature/pre-term rhythmic activity. 
\end{abstract}

\maketitle

\section{Introduction}
Biological phenomena at different length scales often exhibit periodic
activity whose frequency can
span many time scales, and where the rhythmic behavior is
crucial to the functioning of the relevant systems~\cite{Glass01}. 
For instance, in the physiological context, oscillations are observed
in variations of intracellular molecular
concentrations~\cite{Orchard83,Goldbeter97}, activation of pancreatic
$\beta$-cell islets that control the release of
insulin~\cite{Grapengiesser88}, changes in the activity
levels of different brain areas~\cite{Buzsaki04}, circadian rhythms
that control the daily sleep-wake cycle~\cite{Winfree00} and patterns
of locomotion~\cite{Golubitsky99}.
In a system comprising many different oscillating elements, it is
crucial for their activity to be synchronized in order for the
generation of rhythmic activity at the macro- or systems-level.
One of the
most significant biological contexts in which such synchronization is
observed is in the regular electromechanical contraction of the heart.
A specialized group of pacemaker cells located in
the sino-atrial node of the heart coordinates this periodic activity
in a robust manner~\cite{Tsien79,Keener98}. Disruption of the
coherent collective behavior can result in arrhythmia that may be
fatal if untreated in many cases~\cite{Jalife}.
However, such clock-like
centralized coordinating agencies, although seen in several
contexts~\cite{Chigwada06}, cannot be identified for many biological
processes, which raises the possibility that
coherence appears through self-organization among the many interacting
oscillating elements.

A biological system whose function is critically dependent on the
occurrence of rhythmic activity
is the pregnant uterus, where coherent contractions are observed
immediately preceding delivery. However, the available
evidence does not suggest that this is centrally coordinated by a
specialized cluster of pacemaker-like cells, as in the heart. For most
of pregnancy, the uterus does not exhibit any sustained spontaneously
generated excitation and can be well described as an excitable
medium~\cite{Winfree00}.
However, on approaching term, episodes of
periodic activity are observed more and more frequently, and they
increase in magnitude as well as in duration until system-wide powerful
contractions eventually expel the fetus~\cite{Shmygol2007}. 
As the mechanical activity of the uterus is governed by electrical
excitation of the tissue, the rhythmic contractions are related to the
synchronization of oscillations in the transmembrane potentials of the
cells in the myometrium that forms the bulk of uterine wall. 
In pathological
cases (occurring in more than 10\% of all pregnancies), rhythmic
contractions may be initiated much earlier than normal leading to preterm birth
~\cite{Duquette2005}
which account for a third of infant mortality in the
USA~\cite{Popescu2007}.
It is not yet known why in many cases rhythmic activity is initiated
significantly earlier than normal. As there is currently no effective
treatment for events leading to preterm labor, understanding the
dynamical processes underlying the onset of self-organized 
coherent rhythmic activity
in excitable medium may have potential clinical benefits.

The present work is concerned with the emergence of synchronized oscillations
in a system of coupled excitable and passive
cells~\cite{Jacquemet2006,Kryukov2008,Chen2009,Cartwright2000,Petrov:10} 
that 
is motivated by the cellular organization of the mammalian uterus,
where none 
of the constitutive cells can spontaneously oscillate in
isolation~\cite{Shmygol2007, Duquette2005, Young2007}. 
Muscle tissue in the
uterus, like that of many other biological organs, is heterogeneous in
composition with electrically excitable myocytes (smooth muscle cells)
forming the bulk but also containing small fraction of electrically
passive cells like fibroblasts and
telocytes~\cite{Duquette2005,Popescu2007,Young1999}.
The role of inter-cellular communication, including that between
dissimilar cell types, in generating spontaneous periodic activity in
tissue has recently come into focus~\cite{Jacquemet2006,Kryukov2008}.
This is of particular
relevance for the uterus as the
electrical coupling between cells through specialized proteins known
as gap junctions is seen to increase remarkably during pregnancy 
\cite{Miller1989,  Miyoshi1996, Garfield1998}. 

Our model, inspired by the heterogeneous architecture described above,
comprises a lattice of diffusively coupled excitable cells, each of
which may be connected to a number of passive cells. 
As observations made on biological tissue do not reveal any regular
organization in the connection density between excitable and passive
cells~\cite{Popescu2007}, we consider the
number of passive cells connected with an excitable cell to 
be randomly distributed about a given mean density.
Numerical exploration of the model dynamics exhibits a variety of
spatio-temporal regimes at different values of inter-cellular coupling
and
average density of passive cells~\cite{Singh2012}, reminiscent of what
has been observed in
other systems of coupled electrically excitable and passive
cells~\cite{Kryukov2008,Pumir2005,Petrov2009}. 
More generally, our results are relevant for a broad class of systems
that comprise coupled heterogeneous elements, e.g., in the cardiac
context~\cite{Bub2002,Bub2005}.
In qualitative agreement with biological observations of uterine
tissue, it is seen from the model that strong cellular coupling
promotes  synchronized oscillations~\cite{Singh2012}.

While our earlier work provided a global description of the overall
dynamics of the model system~\cite{Singh2012}, using a more detailed approach 
to understand
the genesis of the various dynamical regimes we now point out the
key role played by disorder.
In this paper, we show that the local fluctuation in passive cell
density can result in self-organized emergence of ``pacemaker-like regions",
i.e., clusters of coupled cells that spontaneously oscillate,
eventually activating the entire system. We stress that these are very
different from the specialized pacemaker cells of the heart whose role
cannot be adopted by any of the other cells in cardiac tissue under
normal circumstances, whereas individual members of the
self-organized oscillatory groups referred to above are not 
inherently different from the other cells in the system.
Thus, in the heart, pacemaker function is an intrinsic 
property of certain specialized cells, whereas, in the uterus it is
an outcome of interactions between heterogeneous cell types. 

A given distribution of passive cells attached to each myocyte effectively
represents a particular realization of a quenched disordered system. 
The important role played by
disorder in various phase transitions has been extensively
investigated in physical 
systems~\cite{BinderYoung:86,Mezard_Parisi_Virasoro:87}.
Here, we investigate how the disorder described above influences
the transition to sustained coherent activity in a vital
biological organ. 
Specifically, we consider fluctuations in the macroscopic behavior of
the system, viz., emergence of synchronized oscillations in different
replicas, where a replica refers to a particular realization of 
the disorder. 
If excitable and passive cells are coupled with strength $C_r$, the
combination is capable of oscillating when the number density of
passive cells $f$ coupled to an excitable cell is within a critical
range: 
$f_c^l(C_r) \le f \le f_c^u(C_r)$.
A key aspect of the model is that this
transition between quiescent state and oscillatory
activity occurs through a {\it subcritical} 
bifurcation\cite{Baer:92,Ringkvist:09},
so that the oscillations always have a finite amplitude. 
For $f \lesssim f_c^l$, 
the system exhibits very strong fluctuations
from replica to replica as the coupling between excitable cells $D$ is varied. 
This is because close to $f_c^l$, it is not just the mean value of $f$
over the entire system but also the spatially fluctuating values of
local passive cell density
that are important. The definition of the local density involves
a coarse-graining length, $d$, which is expected to be 
related to the coupling between
excited cells as $ d \propto \sqrt{D}$ on general physical grounds.
Therefore, whether a system oscillates for given values of $C_r$ and $D$
will depend critically on the details of the passive cell
distribution in a specific realization. More specifically, oscillating
groups will emerge from regions where the local passive cell density is
greater than $f_c^l$. The probability of these extreme events, i.e.,
the occurrence of such regions
through statistical fluctuations, increases with system size.
Based on these arguments we derive a scaling relation of the system
dynamics as a function of system size and cellular coupling strengths
that we have verified through numerical simulations.

The paper is organized as follows. In Section II we describe the model
system followed in Section III by a discussion of the properties of an
excitable cell coupled to a variable number of passive cells.
Section IV reports our numerical observations of fluctuations in the
system behavior across different replicas and explains how regions with
high local passive cell densities can act as organizing centers for
oscillatory activity (i.e., ``pacemakers") in the system.
Section V contains the key theoretical result of our paper where we
derive the probability of occurrence of such pacemaker regions as a
function of different system parameters. We show that it has a
logarithmic dependence on system size and obtain a scaling relation.
Finally, in Section VI, we conclude with a summary of 
our results.

\section{The model}
As mentioned above, the system under investigation comprises
electrically excitable as well as passive cells. The dynamics of
excitable cells can be described by equations having the general form
\begin{equation}
C_m \frac{\dd V_e}{\dd t} = -I_{ion} (V_e, g_i) + \Gamma_e,
\label{eq:FHN}
\end{equation}
where $V_e$ (mV) represents the transmembrane potential, $C_m (= 1
\mu$F cm$^{-2}$) is membrane capacitance density, $I_{ion} (\mu$A
cm$^{-2}$) represents the total density of currents transported across
the cell membrane by different ion channels, $g_i$ are variables
describing the gating dynamics of the ion channels and $\Gamma_e$ corresponds to external stimulation that could be due to
coupling with neighboring cells. While the explicit
form for the ionic current density $I_{ion}$ depends on the type of
cell being considered and the level of biological realism desired, in
this article we focus on phenomena that occur independent of the
finer-scale details of specific models 
and therefore use the generic 
FitzHugh-Nagumo (FHN) model~\cite{Keener98}. The functional form for
the ionic current in the FHN system is $I_{ion} = - A V_e (V_e - \alpha)
(1-V_e) + g$, where $A (=3)$ is a parameter governing the fast
activation kinetics, $\alpha (=0.2)$ is the activation threshold 
and
$g$ represents an effective repolarization current. The time-evolution
of $g$ is described by 
\begin{equation}
\frac{\dd g}{\dd t}  = \epsilon (V_e - g), 
\label{eq:slow}
\end{equation}
with $\epsilon (=0.08)$ corresponding to a relatively slow 
rate of recovery of the system from the excited state. 
If the system is capable of oscillation, its period is
governed by the time-scale $\sim
1/\epsilon$~\cite{Keener98}.
For the chosen set of parameter values, the dynamics 
of the FHN system involves either
relaxation towards a fixed point for a subthreshold perturbation or a
finite amplitude excursion before decaying to fixed point after an
interval (action potential) for 
a strong enough or suprathreshold stimulus.
Thus, the active cells are in an excitable regime in which they are
not capable of exhibiting oscillations spontaneously (i.e., in the
absence of external stimulation).

In addition to excitable cells, the system also contains passive cells
described by the dynamics of the transmembrane potential 
$V_p$~\cite{Kohl1994}:
\begin{equation}
\frac{\dd V_p}{\dd t} = K (V_p^R-V_p) + \Gamma_p \,,
\label{eq:passive}
\end{equation}
so that activity of the cell tends to decay at an exponential rate
to its resting value $V_p^R$ (=1.5) with $K$ (=0.25) specifying the
characteristic time scale of relaxation. 
Similar to $\Gamma_e$, $\Gamma_p$ represents the external stimulus to
the passive cell.
When excitable and passive cells are coupled to each other via an
electrical conductance (representing gap junctions in biological
tissue), the external stimulus of each cell corresponds to the input
received from the other cell so that the respective external stimuli 
terms are:
\begin{subequations} 
\label{eq:coupling}
\begin{align}
\Gamma_e = & n_p C_r (V_p - V_e) \,, \\
\Gamma_p = & C_r (V_e - V_p)  \,.
\end{align}
\end{subequations} 
The strength of coupling is represented by $C_r$ while $n_p$
corresponds to the 
number of passive cells coupled to an excitable cell.

To investigate spatially extended patterns that are seen in biological tissue
such as that of the uterus,
we consider a 2-dimensional (2D) system of coupled excitable and passive cells. 
The system comprises a square lattice of $L \times L (=N)$ excitable cells coupled
diffusively to their nearest neighbors on the lattice. The coupling results
in an additional term $D \nabla^2 V_e$ in the term $\Gamma_e$ in Eq.~(\ref{eq:FHN})
where $D$ is the effective diffusion constant (the lattice spacing being set to 1)
and the Laplacian being numerically implemented by a five-point stencil. 

Each excitable cell is coupled to a variable number $n_p$ 
passive cells, with each passive cell being connected to a single 
excitable cell. The total number of passive cells in the system is $M = N \times f$,
$f$ being the passive cell number density. Each passive cell is assigned to
a randomly chosen excitable cell drawn uniformly from the population of $N$ cells.
We use a binomial distribution for $n_p$, which at large $N$ converges to
a Poisson distribution with mean $f$.
Each realization of $M$ passive cells randomly distributed over $N$ excitable cells, 
which is an instance of quenched disorder in the system,
is referred to as a {\em replica}.

We have integrated the system numerically using a fourth-order Runge-Kutta scheme, 
using time step $dt=0.05$. Periodic boundary conditions have been used to
obtain all the results reported here; our earlier results have shown that using
no-flux boundary conditions do not result in qualitatively different phenomena~\cite{Singh2012}. After starting from random initial conditions, the system
is first allowed to evolve for $2 \times 10^4$
time steps (corresponding to transient behaviors) before recording the data. 

\section{Global oscillations and large fluctuations}
\subsection{Single excitable cell coupled to passive cells: The mean-field
limit}

Before proceeding to investigate the spatially extended system, we consider
the dynamical system 
comprising an excitable cell
coupled to one or more passive cells, given by
Eqs.~(\ref{eq:FHN}-\ref{eq:coupling})~\cite{Jacquemet2006,Kryukov2008,Singh2012}.
This problem, which formally corresponds to zero spatial dimension,
can also be viewed as the mean-field limit of extremely strong diffusive
coupling between excitable cells such that spatial fluctuations can
be ignored and the lattice effectively behaves 
as a single excitable cell coupled to the average number of passive cells, $f$.
Even though an excitable cell couples to only an integer number $n_p$ of passive
cells, the mean $f$ will in general not be an integer.
Thus, the integer value of
$n_p$ in Eq.~(\ref{eq:coupling}) can be replaced by a real number $f$, which
corresponds to an exact representation of the system dynamics in the 
mean-field limit. 

\begin{figure*}[tb]
\begin{center}
\includegraphics[width=0.95\linewidth]{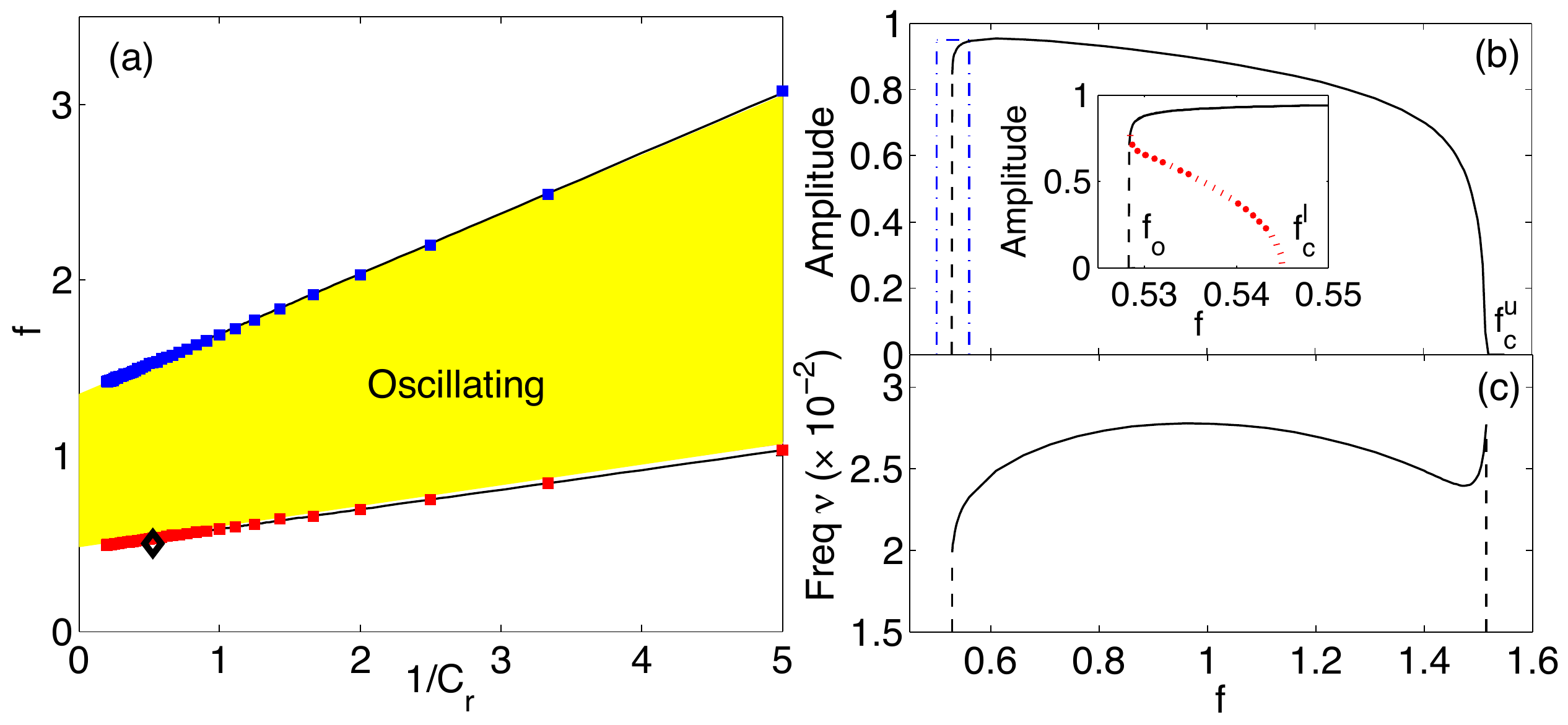}
\end{center}
\caption{ 
(a) Coupling an excitable cell to $f$ passive cells with strength
$C_r$ results in
spontaneous oscillations in the region bounded between the curves
representing $f_c^l (C_r)$ and $f_c^u (C_r)$. 
The shaded region corresponds to the situation where the fixed point
solution is linearly unstable (analytical result)
while the region enclosed by the lines 
indicate where oscillations are observed in numerical simulations.
Most of the 2D results shown in this paper are for $C_r=1.9$ and $f=0.5$,
indicated by the diamond, slightly below 
the critical line $f_c^l$. 
The representation of the ($f,C_r$) space shown here 
demonstrates that the critical
values of $f$ depend linearly on $1/C_r$. The
amplitude (b) and frequency (c) of the
oscillations vary as a function of $f$ (shown for $C_r=1.9$), with
the region enclosed by the dashed-dotted lines in (b) shown in a
magnified view in the inset. Between $f_0=0.529$ and $f_c^l=0.545$ the
system exhibits bistability, which indicates that the Hopf bifurcation
at the lower critical value of $f$ is subcritical. The dotted line
is a schematic representation of the unstable solution.   
}
\label{fig:0d}
\end{figure*}

Fig.~\ref{fig:0d}~(a) shows that the system is capable of exhibiting
spontaneous oscillation over a range of values of $C_r$ and $f$, as
has been reported earlier~\cite{Jacquemet2006}. 
For a given value of coupling $C_r$, the fixed point solution of the
system is unstable when $f_c^l < f < f_c^u$, where the lower and upper 
bounds can be obtained analytically by linear stability analysis of
Eqs.~(\ref{eq:FHN}-\ref{eq:coupling}) around the steady state.
A simple relation between $f_c^{l,u}$ and $C_r$ is obtained in the
limit of large $K$, i.e., when the passive cell relaxes rapidly.
Under this condition Eq.~(\ref{eq:passive}) is replaced by an algebraic
equation such that $\Gamma_e = f K (V_p^R - V_e)/(1 + K/C_r)$. Thus,
as $f$ and $C_r$ appear only via $\Gamma_e$,
the critical value of $f$ depends on $C_r$ as $f_c^{l,u} = c_{l,u} +
m_{l,u}/C_r$, where $c_{l,u}, m_{l,u}$ are independent of $C_r$.
Note that, although the above argument is strictly valid only in the large
$K$ limit, we observe from our numerical results that the 
dependence of the critical values of $f$ on $C_r$ 
follows the above relation even for small $K$ [Fig.~\ref{fig:0d}
~(a)]. Thus, 
$f_c^l(C_r)$ decreases monotonically with increasing $C_r$ and approaches
a finite value as $C_r \rightarrow \infty$.

While the Hopf bifurcation resulting in loss of stability of the fixed point
solution at $f=f_c^u(C_r)$ is supercritical, 
the bifurcation at $f=f_c^l(C_r)$ is subcritical so that
oscillating solutions of {\it finite} amplitude exist for 
$f < f_c^l(C_r)$. For example,
for $C_r=1.9$ we observe oscillations when $f > f_c^l=0.545$ 
[Fig.~\ref{fig:0d}~(b), inset], while they are never observed 
for $f < f_0 \simeq 0.529$. 
Thus, for mean passive cell density in the interval $f_0 < f < f_c^l$,
the system is multistable so that 
both fixed point and oscillatory solutions can be 
observed numerically [see the narrow interval enclosed by the
dash-dotted line in Fig.~\ref{fig:0d}~(b)]. 

In this paper we focus on the situation when $f < f_c^l (C_r)$.
As discussed in detail later for this condition we observe large
fluctuations between replicas when the system is made to undergo 
dynamical transitions by increasing the coupling between excitable
cells, $D$.
In contrast, when $C_r $ is increased so that 
$f > f_c^l(C_r)$,
increasing the coupling $D$
results in a much simpler transition that approaches the mean-field
behavior. 
The replica variations are most pronounced when $f$ just exceeds 
$\lim_{C_r \rightarrow \infty } f_c^l(C_r) $, the minimal 
value of $f_c^l(C_r)$. 
Most of the numerical
results reported here are obtained with $f = 0.5$
[Fig.~\ref{fig:0d}~(a)],
while the minimal value is $f = 0.484$. 

\bigskip

\subsection{Two-dimensional media}

\begin{figure}[tb]
\centering
%\subfloat[] 
%{\includegraphics[width=7.2cm]{model_structure.pdf} }
{\includegraphics[width=0.97\linewidth]{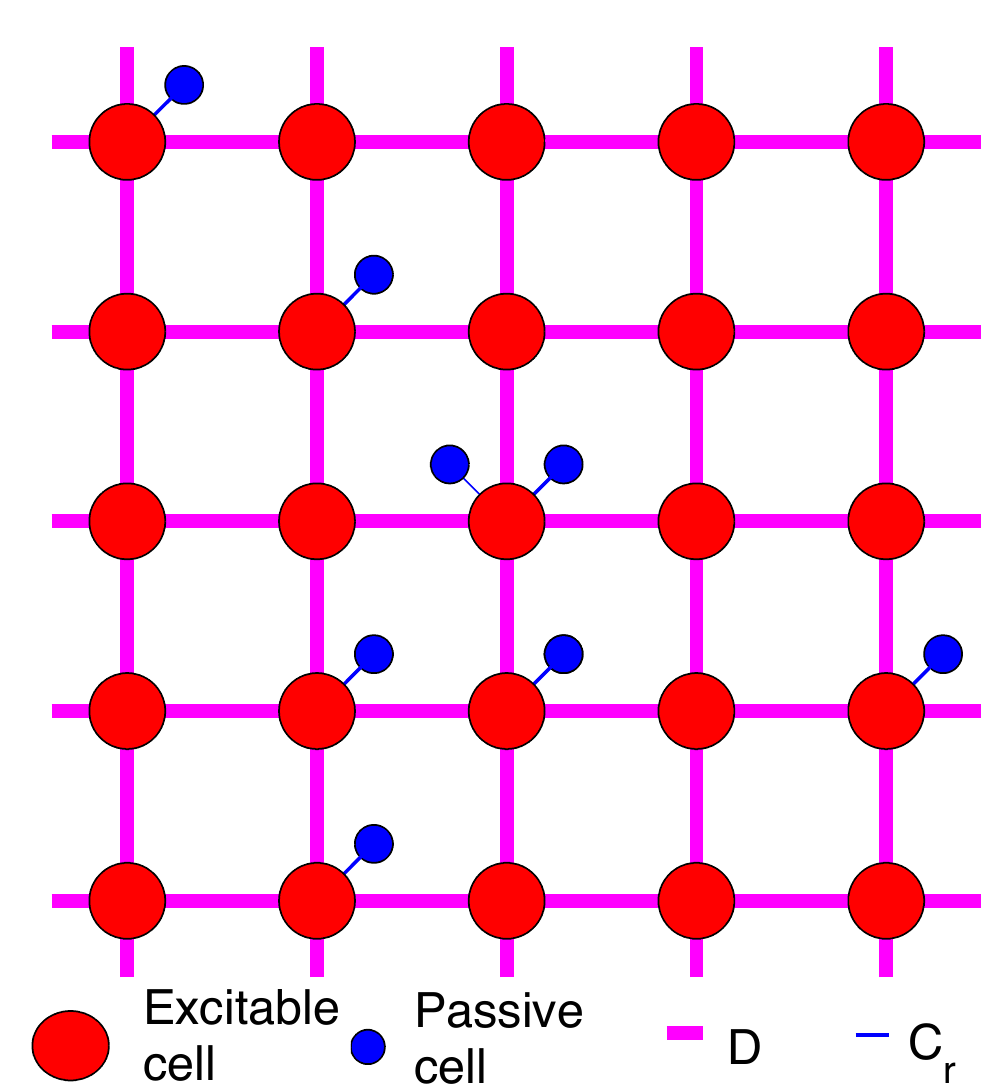} }
%\subfloat[] 
%{\includegraphics[width=7.2cm]{snapshot_frequency_density_pattern_2_13_04.pdf}}
\caption{
Structure of the 2D model system. Excitable cells (large circles)
lie on a regular square lattice, and are connected to their neighbors 
with a coupling strength $D$. In 
addition, passive cells (small circles) are connected to excitable cells 
with coupling strength $C_r$ 
at randomly selected sites of the lattice.
Each choice of the distribution
of passive cells defines a replica.}
\label{fig:def_model}
\end{figure}
\begin{figure*}[tb]
\centering
%\subfloat[] 
%%{\includegraphics[width=7.2cm]{snapshot_frequency_density_pattern_1_13_04.pdf} }
%{\includegraphics[width=7.2cm]{snapshot_frequency_density_pattern_1.pdf} }
{\includegraphics[width=0.49\linewidth]{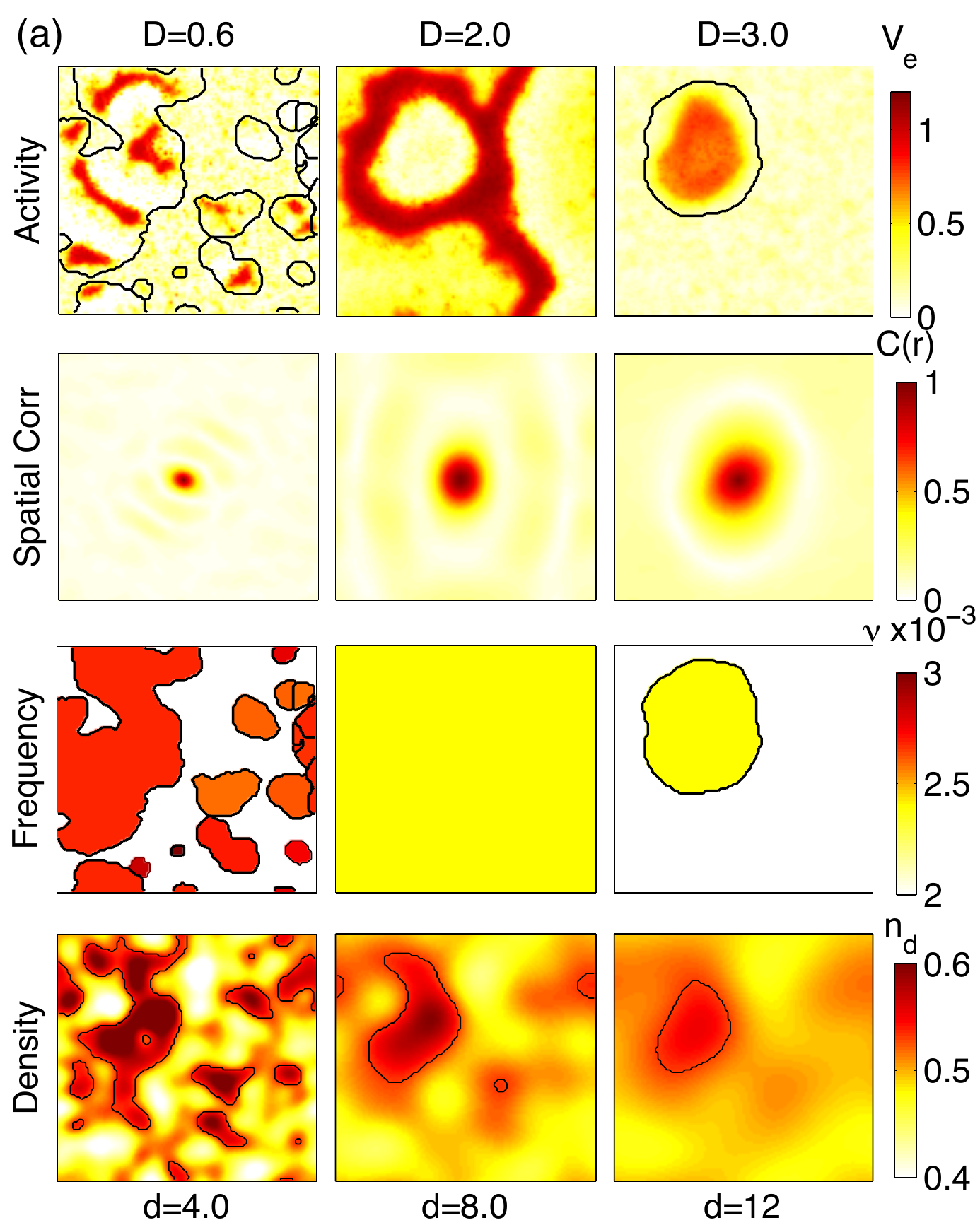} }
%\subfloat[] 
%%{\includegraphics[width=7.2cm]{snapshot_frequency_density_pattern_2_13_04.pdf}}
%{\includegraphics[width=7.2cm]{snapshot_frequency_density_pattern_2.pdf}}
{\includegraphics[width=0.49\linewidth]{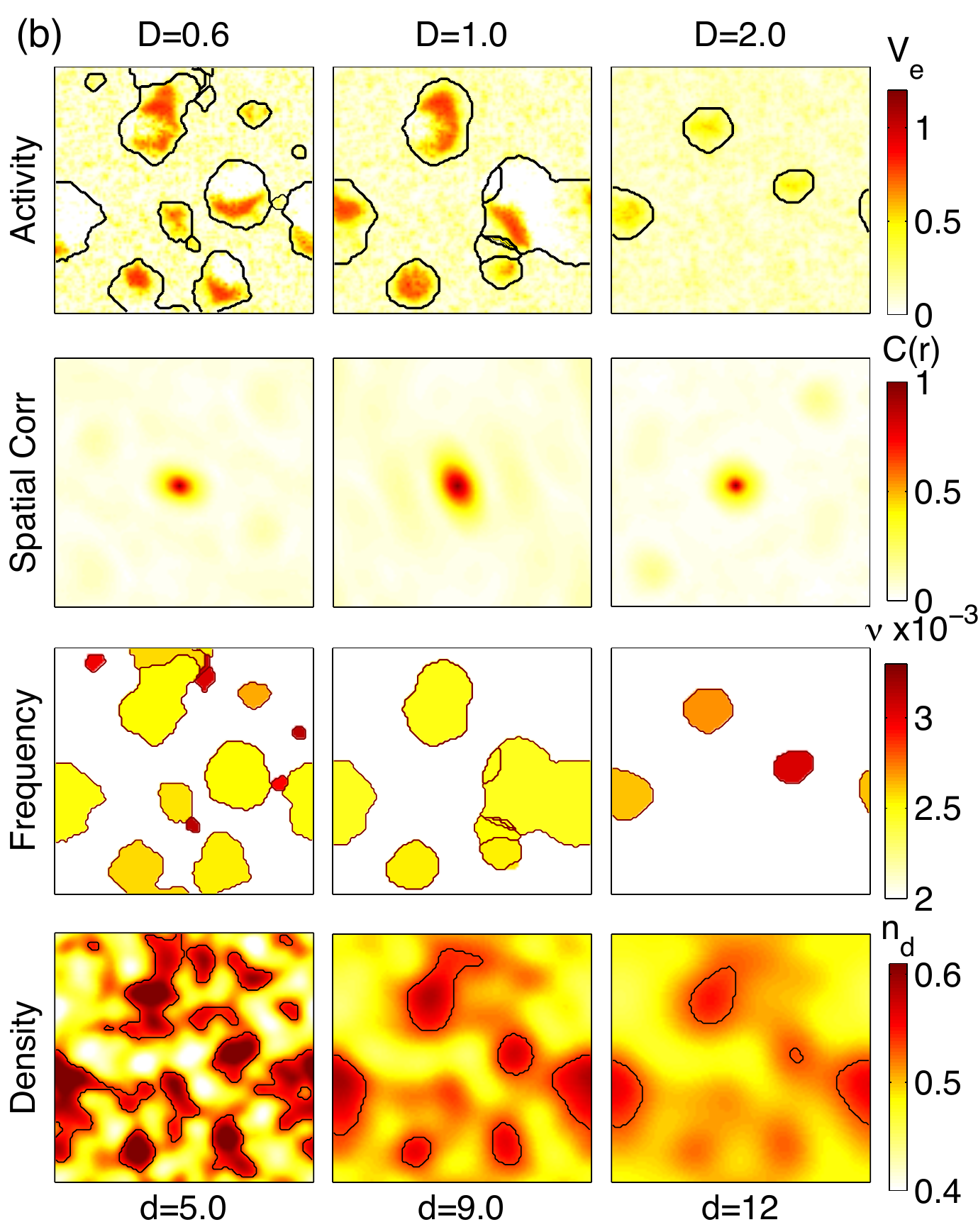}}
\caption{
Replica-dependent fluctuations close to the transition to sustained
oscillations in 2D systems of coupled excitable and passive cells, 
with identical sets of parameters
($f = 0.5$, $C_r = 1.9$, $L = 100$), and increasing values of the coupling
strength $D$.
%Qualitative difference in the the model solution for different replicas. 
Snapshots of activity $V_e$ %(top row) 
%($f = 0.5$, $C_r = 1.9$, $L = 100$) for increasing values of coupling $D$
for two different replicas are shown in the top rows of (a) and (b).
The corresponding time-averaged spatial correlation, 
which provides a characteristic size of the active regions,
is shown in the second row. 
The  frequencies of individual oscillators in the medium
are shown in the pseudocolor plots in the third row (white
corresponds to absence of oscillation).
%The second line shows the frequency averaged over $5000$ time units, 
%corresponding to more than $150$ periods of the system.
The last row shows the local density of passive cells averaged over a length
scale $d$ indicated below each frame. For the replica shown in
(a), we observe global synchronization at $D=2$ followed by
progressive cessation of spontaneous periodic activity in the system 
indicated as a shrinking region of oscillating cells when coupling is increased
further.
%Panel (a) indicates that as $D$ increases, oscillations of all cells with
%the same frequency (global synchronization) is achieved for $D=2$. Increasing
%$D$ to larger values leads to a progressive disappearance of all oscillations; 
%when $D=3$, only one cluster remains, which shrinks when increasing
%further the coupling. \\
However, for the replica in (b), coherent oscillation is not observed
as $D$ is increased, with the existing localized oscillating clusters
having distinct frequencies
gradually decreasing in size. }
%Panel (b) : increasing $D$ does not lead to coherent oscillations: some 
%regions never become oscillatory, and  the clusters finally disappear when
%increasing coupling. \\
%The parameters chosen here are $C_r=1,9$, $f=0.5$ 
%($\mu = 5.7 \times 10^{-2}$); the
%values of $D$ are indicated above each column. 
%The number of cells is $N_x = 100$ ($N = 10^4$).  }
\label{fig:maps1}
\end{figure*}

Biological organs exhibiting electrical activity that are often
functionally important, such as the heart or the uterus, 
are spatially extended objects, being made of tissue comprising a large number 
of excitable and passive cells. As mentioned earlier, we have modeled
these by a 2D system of
coupled excitable and passive cells
with periodic boundaries (Fig.~\ref{fig:def_model}). 
Fig.~\ref{fig:maps1} shows snapshots
of activity in two different replicas of such a system
as the diffusion constant $D$ is increased.
We have investigated the system for a value of the average passive cell
density $ f = 0.5$ and coupling strength between excitable and passive cells  
$C_r = 1.9$.
For these values, 
$f < f_c^l \approx 0.545$, so that in the mean-field limit the medium
does not show any oscillation; however, for finite values of $D$, it
is possible to observe oscillatory behavior either in local
clusters or globally as traveling waves (Fig.~\ref{fig:maps1}).
We define a non-dimensionalized distance $\mu=(f_c^l-f)/f_c^l$
between the system under study and the critical system 
where the fixed point becomes unstable 
in the mean-field limit (for $f=0.5$, 
$\mu \approx 5.7 \times 10^{-2}$).
For small $\mu$, the behavior of 
the system is sensitively dependent on the particular replica chosen
(Fig.~\ref{fig:maps1}). This is manifested as large fluctuations in
the system behavior, measured by various order parameters defined
later. For example, while global synchronization where all cells
oscillate with the same frequency is seen in one replica 
[Fig.~\ref{fig:maps1} (a), $D=2$], another replica for the same set
of parameters
exhibits only localized clusters of cells oscillating with different
frequencies [Fig.~\ref{fig:maps1} (b), $D=2$].

\medskip

\begin{figure}[tb]
\begin{center}
\includegraphics[width=0.99\linewidth]{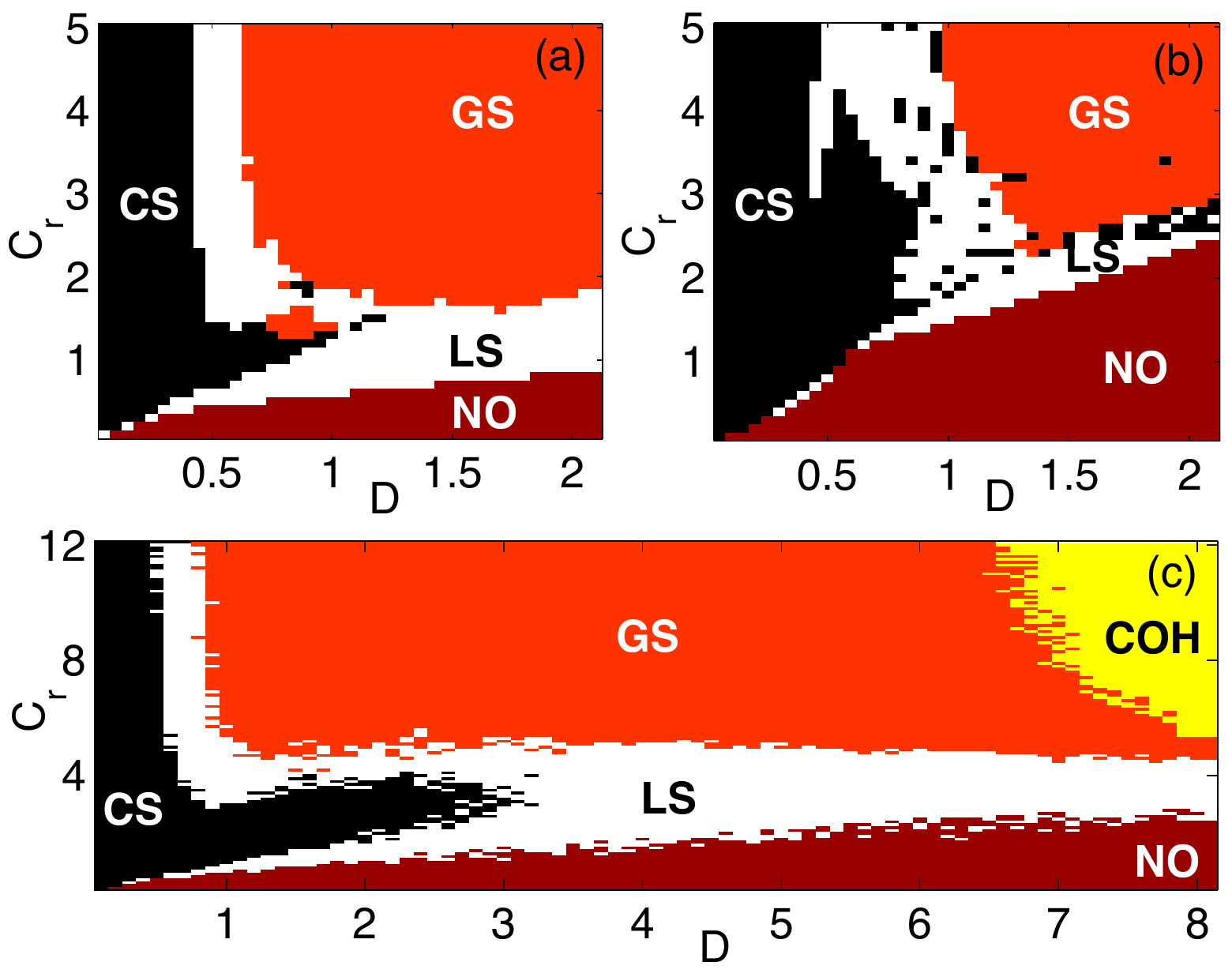}
\end{center}
\caption{
(a-b) Phase diagrams for the 2D system of coupled excitable and
passive cells with $f=0.5$ in two replicas of linear dimensions $L=50$.
Initial values are randomly chosen for each point in the diagram.
%The horizontal broken line corresponds to $C_r$ such that
%$f_c^l (C_r)=0.5$, where a subcritical bifurcation occurs in the
%mean-field limit. Above this line, we will always observe GS at
%large $D$. 
A ``mean" phase diagram obtained by averaging 
over many ($\sim 100$) replicas is
shown in (c) for a system of linear dimension $L=64$. The averaging
implies that if a replica is chosen at random for a given $D$ and
$C_r$, the behavior seen in (c) will be observed with high
probability.}
\label{fig:Phases}
\end{figure}

This sensitive dependence of the system dynamics on the
distribution of $n_p$ can also be seen from the $(D,C_r)$-parameter
space diagrams of the 2D system for different replicas
(Fig.~\ref{fig:Phases}). While for low values of $C_r$, increasing $D$
results in complete absence of oscillations in the system (``No
Oscillation" or NO
phase),
for larger values of $C_r$ we observe clusters of oscillating cells
(``Cluster Synchronization" or CS phase) at
low $D$,
with each cluster having a characteristic frequency that may differ
from other clusters. As $D$ is increased, the clusters gradually
synchronize with each other although isolated regions of
non-oscillating cells can exist (``Local synchronization" or LS phase),
until all cells eventually oscillate at the same frequency at a
sufficiently high $D$ (``Global synchronization" or GS phase). In
numerical simulations, the phases were obtained by using suitable
order parameters (defined in the next subsection), and choosing
appropriate threshold values.
%As can be seen in Fig.~\ref{fig:maps1}, increasing $D$ leads to the
%emergence of global oscillations due to cooperative phenomenon over
%clusters of cells~\cite{Falcke1994,Kanakov:07}. A further increase of
%$D$ leads first to frequency synchronization between locally coupled
%oscillating cells and then to the disappearance of oscillations,
%which we study in detail as a collective behavior.
As can be seen by comparing
Fig.~\ref{fig:Phases}~(a) and (b), the diagrams differ quite
significantly in terms of the actual dynamical regimes that are observed for 
the same set of values of $D$ and $C_r$, illustrating the
significant variability from one replica to the other.
By averaging over many such replicas, we can obtain a ``mean" phase
diagram. The diagram obtained here, Fig.~\ref{fig:Phases}~(c), for $f = 0.5$ 
is qualitatively 
similar
to the one obtained in Ref.~\cite{Singh2012} for a
higher value of $f$ (= 0.7). The region corresponding to Coherence
or ``COH" phase also exists in the present case, appearing at large
value of $D$ which is outside the region of interest of the present paper.
%By comparison, the differences in the observed behavior 
%when $f = 0.5$, in the case studied in this article, are much stronger 
%than when $f = 0.7$. As hinted above, we relate this feature of 
%the present system to the fact that $f$ is close to the transition
%point $f_c^l$ (Fig.~\ref{fig:0d}). 

\medskip

\subsection{Order parameters}
For a detailed quantitative analysis of the spatiotemporal dynamics of
the 2D system, we have used two order parameters: (i) the fraction of
oscillating cells in the medium $f_{osc}$ and (ii) the width of the
frequency distribution as measured by
the standard deviation $\sigma_{\nu}$, where $\nu$ denotes the
frequencies of the oscillating cells. 
The amplitude of an oscillating cell is obtained from the
%$A(i,j)$ of the oscillations
%at site $(i,j)$ as the 
square root of the integral of power spectral density (PSD)
of the corresponding $V_e$ time-series, $\nu$ being the frequency
($>0$) at
which the PSD is maximum. 
%This definition accurately 
%accounts for non-monochromatic oscillations.
The fraction of oscillating cells in the system $f_{osc} = N_{osc} / N$ 
is
the ratio of the number of oscillating cells 
$N_{\rm osc}$, i.e., those having amplitude higher than a chosen
threshold, to the total number of cells $N$.
%\begin{equation}
%\label{eq:def_nosc1}
%f_{\rm osc}=\frac{N_{\rm osc}}{N} % = \frac{1}{N}\sum_{(i,j) \in {\mathbb I}} H(A(i,j)-A_{\rm thr}) = mean( H(A - A_{thr} ) ) \,,
%\end{equation} 
%In practice, the number $N_{\rm osc}$ is measured by counting oscillating cells with an amplitude larger then a threshold value $A_{\rm thr}$, 
%The threshold is typically chosen to be 10\% of the maximum amplitude).
%The function $H$ is the standard Heaviside function, equal to $1$ ($0$) for positive (negative) values of its argument.
Note that the oscillating cells have approximately the same 
amplitude which is a consequence of the subcritical nature of 
the transition across $f_c^l$ (Fig.\ref{fig:0d}). This enables a clear
distinction between oscillating and non-oscillating cells, so that the
value of $f_{osc}$ does not depend sensitively on the choice of the
threshold.
%The amount of activity can also be quantified by the averaged amplitude
%of the oscillations over the all system, defined as:
%\begin{equation}
%\label{eq:def_Am}
%A_m = \frac{1}{N} \sum_{(i,j) \in {\mathbb I}} A(i,j) = mean(A) \,.
%\end{equation}

\begin{figure*}[tbp]
\begin{center}
\includegraphics[width=0.98\linewidth]{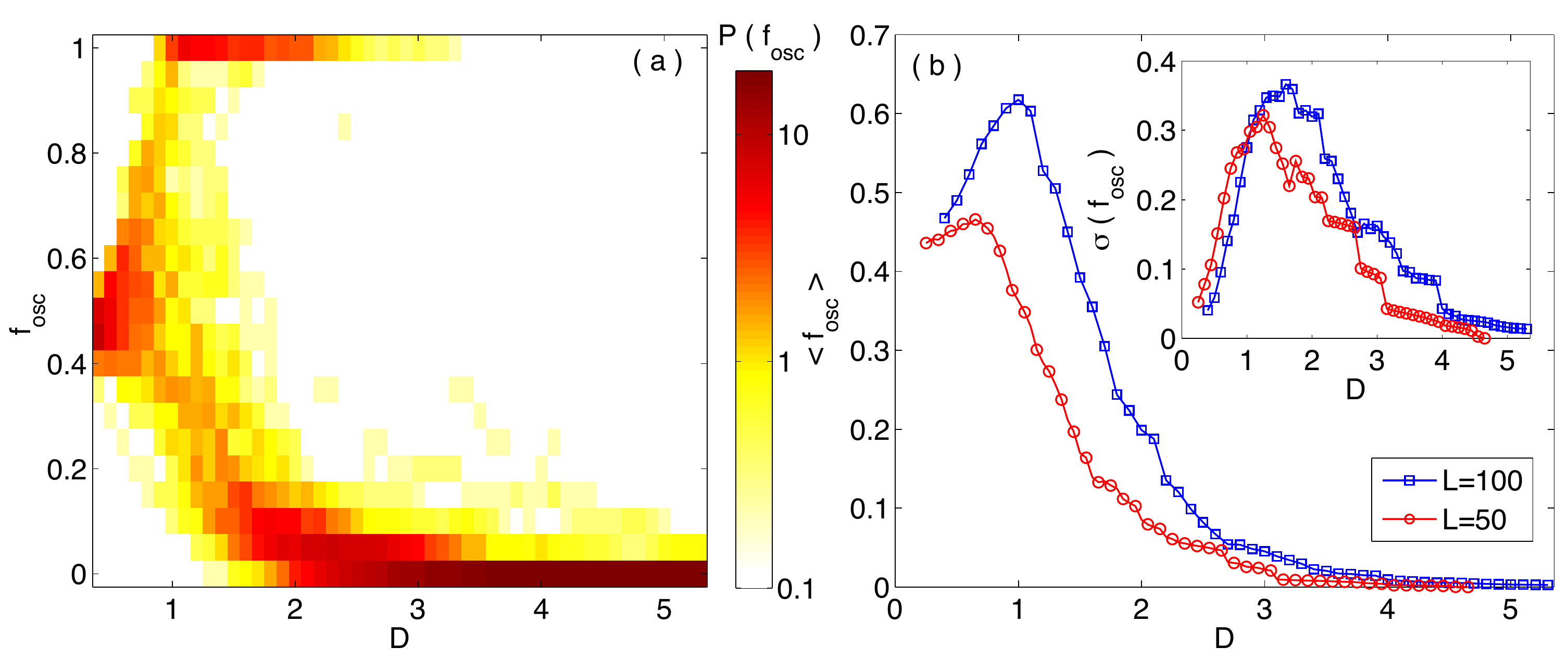}
\end{center}
\caption{
(a) Probability distribution function of the fraction of oscillating cells 
constructed from $176$
different replicas of a 2D 
system of linear dimension $ L = 100$, with identical values of $f = 0.5$ 
($\mu = 5.7 \times 10^{-2}$),
and $C_r = 1.9$, shown as a function of diffusive coupling strength $D$. 
(b) Variation of the mean value of $f_{osc}$, and that of the standard deviation
(inset) with $D$, both for systems of size 
$L = 100$ (squares) and $L = 50$ (circles).
Starting from the uncoupled case $D = 0$, as the coupling strength becomes 
larger
we observe an increase
in the fraction of oscillating cells in the system, in certain cases
resulting in 
global synchronization. Further increase of the coupling eventually
leads to complete cessation of any activity in the system. The fluctuation 
about the mean is maximum at $D \lesssim 2$ (inset), when the distribution
exhibits a strong bimodal nature.
}
\label{fig:histo}
\label{fig:A_Nosc}
\end{figure*}

%The fraction of oscillating cells $f_{\rm osc}$
%depends on the replica under consideration.
The different phases shown in Fig.~\ref{fig:Phases} are defined in
terms of the two order parameters as follows. The LS and GS phases
both have $\sigma_{\nu} \rightarrow 0$; however, for the former
$f_{osc} < 1$ while for the latter $f_{osc} \simeq 1$. The CS phase
has a finite $\sigma_{\nu}$ while the NO phase is characterized by
$f_{osc} \simeq 0$. 
The order parameters for a given system depend on the
exact form of the quenched spatial disorder and we can investigate 
the statistical properties of
distribution of $f_{osc}$ 
over an ensemble of replicas (Fig.~\ref{fig:histo}~a). At very low values of
$D$, the distribution of $f_{osc}$ is peaked about a value close to 
$f_{osc}  = f$.
Upon increasing $D$,
the distribution broadens with a bias towards large values of
$f_{osc}$. A peak at $f_{osc}  = 1$ develops around $D \lesssim 1$. 
Increasing $D$ further, one observes a higher probability for very 
small values of $f_{osc}$. At $D \lesssim 2$, the distribution has an 
almost bimodal form, corresponding to realizations with either very
few oscillating cells, $f_{osc} \ll 1$, or almost all cells oscillating 
$f_{osc} \approx 1$. At yet larger values of $D$, the peak at $f_{osc} \approx 1$
disappears, and the distribution concentrates
around $f_{osc} = 0$, implying a strong reduction 
in spontaneous activity of the system. This is
reflected in
Fig.~\ref{fig:A_Nosc}~(b), which shows the ensemble average $\langle f_{osc}
\rangle$  (where $\langle  \rangle$ denotes an averaging 
over replicas) and the fluctuations
about the mean (inset) as a function of the coupling $D$ 
for 2D systems with $L = 50$ and $100$ at $C_r = 1.9$.
We observe that at large values of $D$, the fraction of oscillating cells
in the system decreases on average with $D$, and hence is consistent with
the mean-field result. The standard deviation of $f_{osc}$ becomes
extremely high in the range $1 < D < 2$ which is a signature of the large
fluctuations between different replicas.
Comparing between the two curves for $L = 50$ and 100 shows
that the larger system has a higher probability of showing
oscillations than the smaller one for the same parameter values.
This dependence on the size of the medium is in fact systematic and
is discussed in section~\ref{sec:stats}.

%\medskip

\medskip

%%%%%%%%%%%%%%%%%%%%%%%%%%%%%%%%%%%%%%%%%%%%%%%%%%%%%%%%%%
\section{Fluctuations of passive cell density and local activity}

The behavior described in the previous section can be qualitatively
understood by noticing that diffusion effectively couples excitable
cells that are within a distance $\propto \sqrt{D}$ of each other. 
As a consequence, one can expect that the behavior of a given
excitable cell depends on the density of passive cells in its
neighborhood characterized by the distance $\sqrt{D}$.
%This suggests to relate the observed activity of a
%cell in the system to the fluctuations of the density of passive cells
%in a domain surrounding the cell. 
With this motivation, we begin by describing the procedure for
computing 
the local density of passive cells surrounding a given excitable cell.

\subsection{Averaging procedure}
\label{sec:averaging}
We define local passive cell density $ {\bar n }_d$ coarse-grained over a 
length scale $d$ as the convolution 
$ {\bar n }_d = K_d*n_p$
of the spatial distribution
of passive cells $n_p(i,j)$ ($i,j = 1, \ldots, L$) 
over the lattice with a 2D averaging kernel 
$K_d(i,j)$. For simplicity we have
considered separable kernels, i.e., $K_d(i,j) = k_d(i/d) \times k_d(j/d)$. 
As the coupling term is effectively described by a Laplacian, 
a natural choice for the coarse-graining kernel $K_d$ is the Gaussian kernel
$K^G_d$ given by:
\begin{equation} 
k^G_d(i) = \frac{1}{\sqrt{2\pi d^2}}\exp(-\frac{i^2}{2 d^2}) \,,
\label{eq:def:kernel:Gaussian}
\end{equation}
which has been used for the numerical investigation of the model system
presented in this section.
For analytical convenience we have also used the
simpler ``square top hat" kernel, $K^{STH}_d$ in section
\ref{sec:stats} which is based on the classical 1D top hat filter:
\begin{equation} 
k^{TH}_d(i) = \frac{1}{d} H( d/2-|i| ) \,,
\label{eq:def:kernel:square}
\end{equation}
where $H(.)$ is the Heaviside step function, i.e. $H(x) = 0 $ when $x \le 0$ 
and $H(x) = 1$ otherwise.
For the square top hat kernel, ${\bar n}_d(i,j)$ is the number of passive cells,
averaged over a square subregion of size $N_d=d^2$ centered at the point 
$(i,j)$. In the case of the Gaussian kernel $K^G_d$, the contribution
from a site $(i',j')$ to $n_p$ at site $(i,j)$ depends on the distance between 
the two.
The variance of ${\bar n}_d$ can be written 
as $\sigma_d^2=\sigma^2 / N_d$, where $\sigma^2 $ is the variance of 
$n_p$, and $N_d$ ($ = 4 \pi d^2$) is the effective number of sites 
contributing 
to ${\bar n}_d$. 
%We define the effective number of points $N_d=4\pi d^2$ that
%contributes significantly to a particular site by the condition
%that the standard deviation $\sigma_d$ of ${\bar n}_d$ is related to the variance
%$\sigma^2$ of $n_p$ by 
%$\sigma_d^2=\sigma^2 / N_d$. 

\subsection{Emergence of pacemaker-like regions through diffusion}
\label{subsec:emergence}
%The third line of Fig.~\ref{fig:maps1} shows how the passive cell density 
%$\bar{n}_d$ evolves when $d$ is increased. The patterns of $\bar{n}_d$ and 
%of the oscillation amplitude are clearly similar. 
As can be seen from
Fig.~\ref{fig:maps1}, large-amplitude oscillations occur in regions
that have the highest local density of passive cells.
In addition, the features observed on increasing
the diffusion coefficient $D$ resemble the patterns of the local
passive cell density seen upon increasing the averaging size $d$.
%Indeed, increasing $D$ leads to a reduced number of oscillating cells 
%as a larger diffusion coefficient leads to an effective averaging over
%neighboring cells. 
As $f < f_c^l$, increasing $D$ is expected to ultimately 
suppress oscillation in Fig.~\ref{fig:maps1}
in agreement with the mean-field analysis. 
Indeed, we observe that at larger values of $D$, the number of
oscillating cells is reduced on average.
The coarse-graining procedure reflects this phenomenon as with
increased kernel width $d$ the variations in $\bar{n}_d$ are reduced
significantly. This decreases the probability that a cell will 
have sufficiently high local passive
cell density to generate oscillations.

%To go beyond the qualitative impression given by Fig.~\ref{fig:maps1},
We define a ``pacemaker-like region" to be a group of adjacent cells 
with a local passive cell density $\bar{n}_d$ larger than the threshold
$f_c^l(C_r)$. 
The oscillatory activity arising in these regions 
may propagate in the form of waves to the rest of the system, so that they 
effectively act as
% ``pacemaker-like regions'' 
pacemakers that are seen in 
systems having specialized coordination centers.
%which may pro oscillations emerge in the form of waves, this region
%appears to initiate such activity thereby resembling pacemakers seen in
%systems with specific coordination centers.
%If there are several such regions,  the case of multiple  Fig.~\ref{fig:0d}(c), close to the threshold, the higher 
%the density, the higher the frequency, which implies that pacemaker-like
%regions are entraining the system, and setting the pace of the oscillations. }
%[assuming it is $> f_c^l(C_r)$].
%When system-wide oscillations emerge in the form of waves, this region
%appears to initiate such activity thereby resembling pacemakers seen in
%systems with specific coordination centers.
%The analysis carried out for a single cell ($0$-d) suggests that the
%cells in a pacemaker-like region are the ones that will lead to
%oscillation.

The number of cells comprising the pacemaker-like region varies from
replica to replica. When $f < f_c^l$, we expect the size of the region
to shrink with increasing coarse-graining length $d$, eventually
disappearing at large $d$ (as predicted by mean-field analysis).
Thus, for a given replica, we can measure the largest value of the
coarse-graining length, $d^*$, for which a pacemaker-like region
still exists.
This is the smallest value of $d$ for which $\bar{n}_d < f_c^l$ 
everywhere in the 2D system. 

\begin{figure}[tb]
\begin{center}
\includegraphics[width=0.98\linewidth]{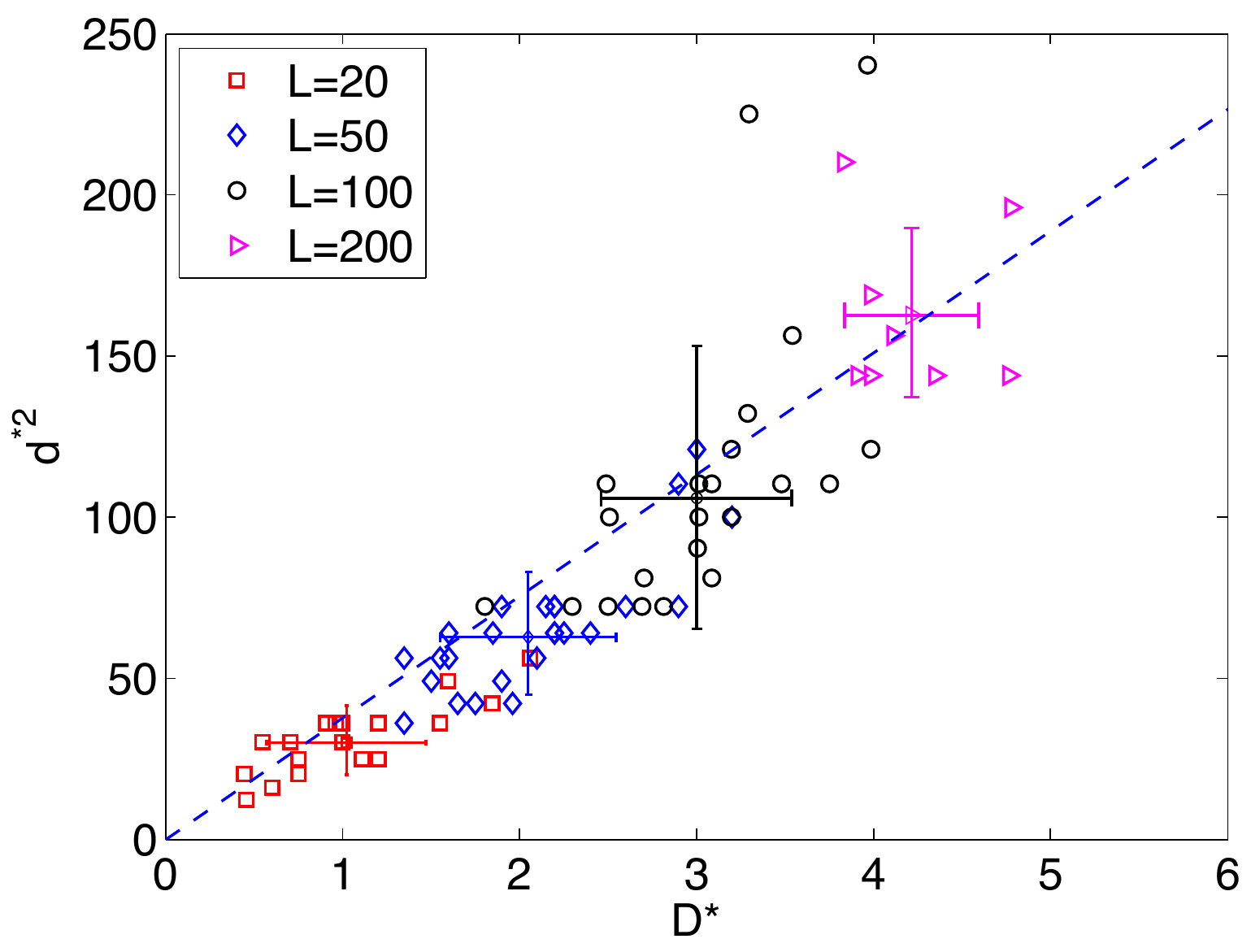}
\end{center}
\caption{ 
%{\bf (Use $D^*$ and $d^*$ instead of $D$ and $d$. Replace
%$N_x$ with $L$ in legend.)}
Relation between the largest values of coarse-graining length $d^*$
and diffusion coefficient $D^*$ for which a given 2D system possesses regions 
with oscillatory activity.
Each point corresponds to a different replica whereas different symbols
represent different system sizes $N=L^2$. Error bars
centered around averages of replicas for the same system size
express the standard deviation. The linear fit between $d^{*2}$ and 
$D^*$, shown as a broken line, has a
slope $T \simeq 36.6$ that lies in the middle of the range of observed
oscillation periods (lying between $31$ and $40$) for different values of the
diffusive coupling strength in a 2D
system with $L=50$.
}
\label{fig:D_d}
\end{figure}

As mentioned above, the coarse-graining length $d$ is related to the
diffusion coefficient $D$, expected to be of the form $d \sim
\sqrt{D}$. Therefore, in analogy with $d^*$, 
we can define a value $D^*$ of the diffusion coefficient
above which the number of oscillating cells in the system goes to zero.
Fig.~\ref{fig:D_d} demonstrates that the relation between $d^*$ and
$D^*$ (shown for different replicas and system sizes) follows
$d^{*2} \approx T D^*$, where $T$ is the slope of the linear fit and
defines a characteristic time of the order of the slow time 
scale in the system, $\sim 1/\epsilon$.
%This relation between the diffusion 
%coefficient $D$ and the averaging size $d$ will be useful in the analysis 
%of Section~\ref{sec:stats}.
%In fact,
%the measured value, $T=36.6$ is in the middle of the range of observed
%period of oscillations, which from $40$ down to $31$ when $D$ is
%increased in a
%$50 \times 50$ system). 
%We do not have any convincing explanation relating
%$T$ to the oscillation period. 
We note that as system size increases, both $D^*$ and
$d^*$ increase. Thus, for systems just below the critical threshold
for oscillatory activity, i.e., $f \le f_c^l$, oscillations are seen
for a wider range of $D$ in larger systems.
%larger systems in the subcritical region $f \le f_c^l$, oscillations
%exist over a larger range of $D$. 
This size dependence is
investigated quantitatively in section~\ref{sec:stats}.

%%%%%%%%%%%%%%%%%%%%%%%%%%%%%%%%%%%%%%%%%%%%%%%%%%%%%%%%%%
\section{Statistical description}
\label{sec:stats}
%The relation between clusters of oscillating cells and 
%pacemaker-like regions, suggested by Fig.~\ref{fig:maps1} is confirmed
%by the proportionality between the values of $D^*$ and $d^*$ seen in
%Fig.~\ref{fig:D_d}. Therefore the fluctuations observed when 
%considering different replicas can be simply understood by investigating
%the presence of pacemaker-like regions as a function of the coarse-graining
%size $d$.
In this section, we investigate the effect of coarse-graining length
$d$ on the presence and spatial extent of pacemaker-like regions in
the 2D system.
%In this spirit, in order to understand the main features observed numerically,
%we analyze here the effect of coarse-graining on the existence and 
%spatial extension of pacemaker-like regions. 
For analytical convenience, we use the `square top hat' kernel,
Eq.~(\ref{eq:def:kernel:square}), for coarse-graining.

\subsection{Distribution of passive cells}
Consider a 2D lattice with $N = L^2$ excitable cells in which a 
%If in a system comprising a 2-dimensional lattice of $N=N_x^2$
%excitable cells a 
total number of $M=f ~ N$ passive cells are
randomly distributed.
The probability that there are $N_p$ passive cells in a region
containing $N_d=\zeta N$ ($0 < \zeta < 1$) excitable cells
is given by the binomial distribution:
\begin{equation}
p^N_{N_d} (N_p) = \binom{M}{N_p} \zeta^{N_p} (1-\zeta)^{M-N_p}.
\label{eq:binom}
\end{equation}
When $N\rightarrow \infty$, this reduces to a Poisson distribution
with mean $f N_d$ : 
$$
p^\infty_{N_d} (N_p) = \frac{(f N_d)^{N_p}e^{-f N_d}}{N_p!}.
$$
%Averaging according to the procedure outlined in 
%section~\ref{sec:averaging} 
For a square top hat kernel, the averaging occurs over $N_d=d^2$ sites where
$d$ is the coarse-graining length and the local passive cell density
${\bar n}_d = N_p/N_d$.
For a Gaussian kernel, the corresponding coarse-grained region
comprises $N_d=4\pi d^2$ sites. 
%{\bf What does the following sentence mean: For both 
%kernels, $N_d$ is given by the dependance of the variance of the random 
%variable $n_d$ in $d$ : $\sigma_d^2 = \sigma^2 /N_d$}.

\subsection{Probability of the occurrence of a spontaneously oscillating cell
in a neighborhood of $N_d$ cells}
\label{subsec:5.2}
\begin{figure}[tb]
\begin{center}
\includegraphics[width=0.99\linewidth]{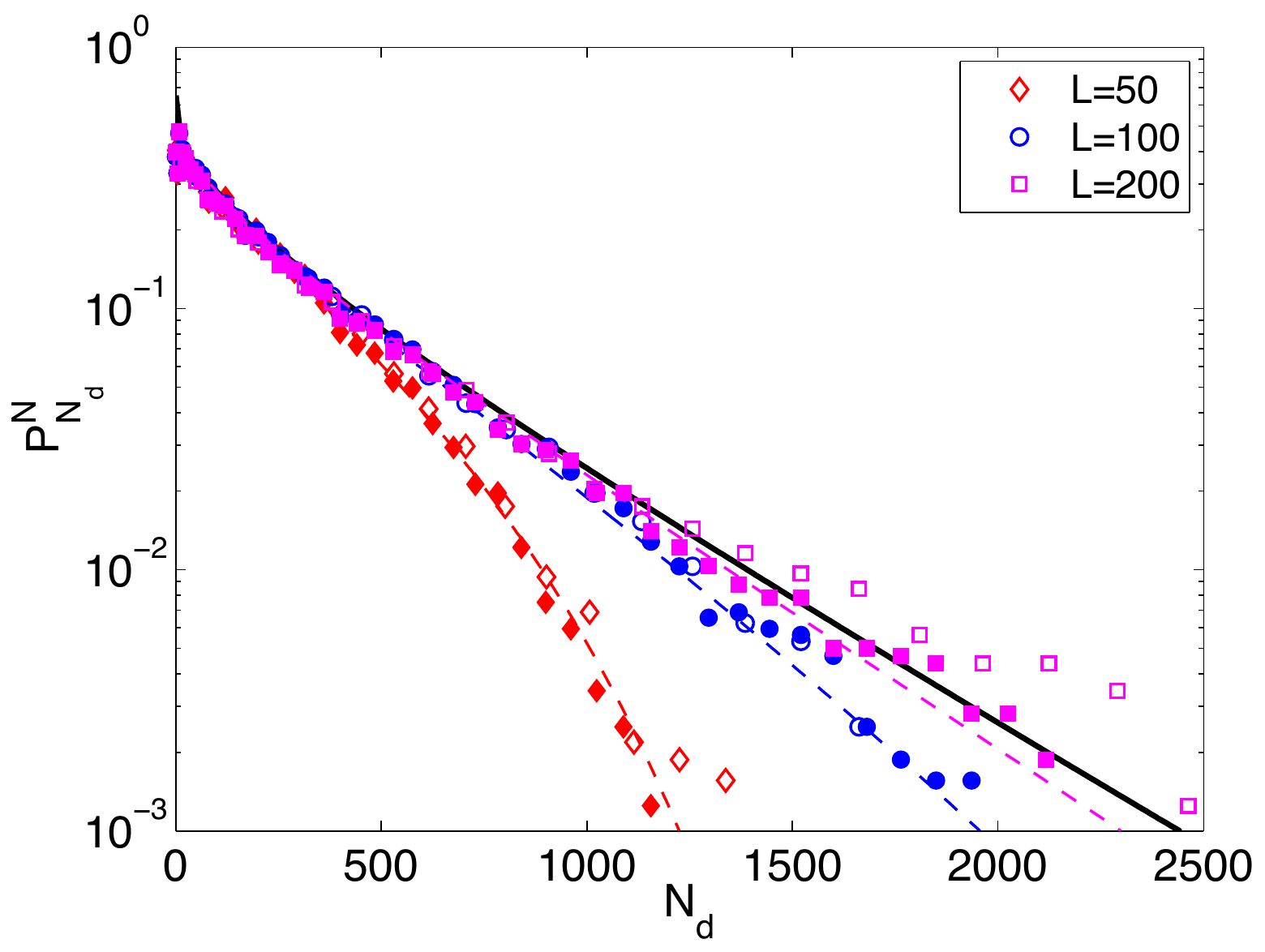}
\end{center}
\caption{
Probability $P^N_{N_d}$ of having local passive cell
density greater than $f_c^l$ in a neighborhood of 
$N_d$ excitable cells as a function of $N_d$. Open (filled)
symbols represent numerical observations using Gaussian (square top hat) 
kernel. 
The broken curves indicate the corresponding probabilities obtained
from the analytical expression of $P^N_{N_d}$, Eq.~(\ref{eq:PNd_finite}). 
The solid curve shows $P^\infty_{N_d}$, the probability values obtained for
infinitely large system, Eq.~(\ref{eq:PNd_infinite}).}
%4 subsets per replica have been used to compute the probabilities
%numerically, which may explain the discrepancy for $N_x=200$
\label{fig:P_pacemaker}
\end{figure}

%group of $N_d$ cells being a pacemaker-like region
Given the probability distribution of passive cells, we now determine
the probability that a given cell is capable 
of spontaneous oscillations when it is effectively coupled through
diffusion to a neighborhood consisting of $N_d$ cells.
This is obtained by considering the probability that the
local passive cell density
$\bar{n}_d = N_p/N_d \ge f_c^l$, which is given by the cumulative
distribution corresponding to Eq.~(\ref{eq:binom}):
\begin{equation}
P^N_{N_d} = \sum_{N_p = f_c^l N_d}^M p^N_{N_d}(N_p) = I_\zeta(f_c^l
N_d, M - f_c^l N_d +1),
\label{eq:PNd_finite}
\end{equation}
where $I_x(a,b)$ is the regularized (incomplete) beta 
function~\cite{Cramer:57} and $P^N_{N_d}$ depends on system size $N$.
In the limit ${N\rightarrow \infty}$, $P^{N}_{N_d}$, tends to a regularized
incomplete gamma function \cite{Abramowitz:72},
\begin{equation}
P^\infty_{N_d} = \frac{\gamma(f_c^l N_d, f N_d)}{\Gamma(f_c^l N_d)}.
\label{eq:PNd_infinite}
\end{equation}
The probabilities obtained for different values of $N_d$ 
from the above expression agrees well with the corresponding
values numerically obtained from different replicas of the 2D system
in Fig.~\ref{fig:P_pacemaker}. 
%shows a comparison between the numerical 
%measurements of the probabilities and the above expressions. 
%Although Eqs.~(\ref{eq:PNd_finite}-\ref{eq:PNd_infinite}) are
%obtained using a top hat square kernel assuming a simple local
%average as the one given by the convolution with the square kernel
%Eq.~(\ref{eq:def:kernel:square}), they also represent perfectly the
%measurements of the passive cell density obtained by convolution with
%the Gaussian kernel Eq.~(\ref{eq:def:kernel:Gaussian}), as can be
%seen in Fig.~\ref{fig:P_pacemaker}.

In the limit of large system size $N$, $P^{\infty}_{N_d}$ can be rewritten by
expressing the incomplete beta function in Eq.~(\ref{eq:PNd_finite})
as a continuous fraction \cite{Abramowitz:72} and keeping only lower
order terms in $N_d^{-1}$ and $N^{-1}$. In addition, 
neglecting $\zeta = N_d/N$ we obtain for any fixed value of $\mu$:
\begin{equation}
P^\infty_{N_d} \simeq \frac{1}{\sqrt{2\pi f_c^l N_d}}\frac{{\rm e}^{-\lambda f_c^l N_d}}{\mu} \,,
\label{eq:P_Nd_simple}
\end{equation}
where $\lambda=-\mu-\log(1-\mu)$. This expression can also be obtained
from Eq.~(\ref{eq:PNd_infinite}) by expanding the regularized 
incomplete gamma function and using the Stirling approximation. 
%{\Q {\bf Do we still need to introduce the variable $\chi$}?}
For small $\mu$, we have $\lambda \simeq \mu^2/2 > 0$ and on introducing a 
reduced variable ${\cal X}=\mu^2 f_c^l N_d$ one obtains:
\begin{equation}
P^\infty_{N_d} \simeq \frac{{\rm e}^{-{\cal X}/2}}{\sqrt{2\pi {\cal X}}}  \,.
\label{eq:X}
\end{equation}

Note that in order to ensure that all replicas for a given value of
$f$ have exactly the same mean number of passive cells $f = \bar{n}_p$
connected to an excitable cell, 
we have used a binomial distribution of $n_p$ for generating the different 
replicas in our numerical study.
%with our numerical results  choice of the distribution function for the number of
%passive cells coupled to each excitable
%cell in the system is dictated by the constraint 
%of having a constant mean fraction $f =\bar{n}_p$, the averaging being 
%over all cells for a given replica. 
%We use the binomial distribution
%for this purpose, as 
In comparison, the Poisson distribution used in our analysis leads
to a mean fraction of passive cells attached to an excitable cell that 
varies from one replica to another for a given value of $f$.
In the limit of 
large system size, the fluctuations in the number of passive cells
become very small, so that both binomial and Poisson distributions lead 
to the same behavior. 
However, when $N$ is not very large,
these fluctuations 
are appreciable when using the Poisson distribution which introduces biases in
the results (see Figs.~\ref{fig:P_pacemaker} and \ref{fig:P_1}).
In addition, for finite $N$, corrections of order $N_d/N$ are expected in the
expression involving the reduced variable $ \cal{X} $, Eq.~(\ref{eq:X}).
%account , while in the $N\rightarrow \infty$, the properties of the system are 
%described in terms of ${\cal X} \propto N_d \mu^2$, corrections of the order
%$N_d/N$ are to be expected when $N$ is finite.
%This modifies our 
%predictions and the scaling in ${\cal X}=D\mu^2$ has to be corrected by 
%terms of the order $\zeta = N_d/N = 4\pi TD/N$. 
These effects are responsible for deviations between 
the predictions
of our analysis and the numerical results.

\subsection{Probability of the occurrence of a pacemaker-like region in the 2D system}

\begin{figure}[tb]
\begin{center}
\includegraphics[width=0.99\linewidth]{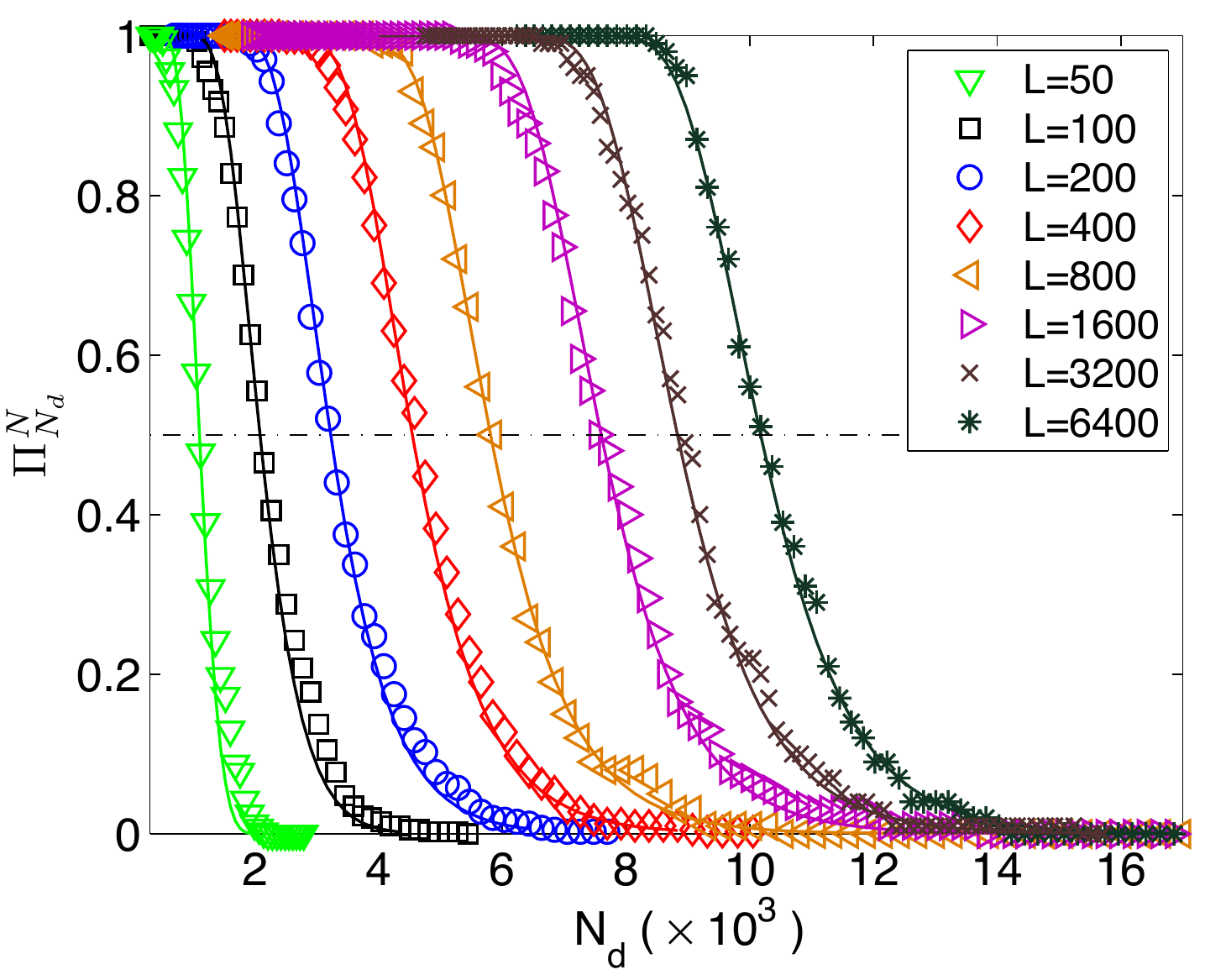}
\end{center}
\caption{
Probability $\Pi^N_{N_d}$ of having a pacemaker-like region in a 2D system of size $N = L^2$ as a function of the coarse-graining size 
$N_d$. Different symbols represent different systems sizes from 
$L  =  50$ to 6400. The continuous lines indicate 
fits to the theoretical expression for $\Pi^N_{N_d}$ given by Eq.~(\ref{eq:P_1_pace}).}
\label{fig:proba_at_least_1}
\end{figure}

As mentioned earlier, Fig.~\ref{fig:D_d} shows $d^*$, the largest value of the coarse-graining
length $d$ for which a pacemaker-like region exists in the 2D system, as a function of the diffusion
coefficient.
%For a single realization/replica, we measured $d^*$ (plotted in 
%Fig.~\ref{fig:D_d}) as the largest value of $d$ for which there is a 
%pacemaker-like region in the system. Ultimately, the last pacemaker-like region observed when increasing the diffusion %coefficient would be composed of a single cell.
We define the probability $\Pi^{N}_{N_d}$ of having a pacemaker-like region in the system as the probability 
of having at least one cell with $\bar{n}_d \ge f_c^l$ in a replica of size $N = L^2$ cells when 
the coarse-graining is done over a neighborhood of $N_d$ cells. 
%From the numerical point of view, it is computed 
%as the average over $400$ realizations of the step function that associates to 
%$N_d=4\pi d^2$ the value 
%
%\begin{equation}
%\begin{split}
%& \quad 1\quad{\rm if} \quad \max_{\mathbb I}(\bar{n}_d(i,j))\ge f_c^l ,\\
%& \quad0 \quad {\rm otherwise} 
%\end{split}
%\label{eq:def:Pi}
%\end{equation}
%A single step function vanishes for $N_d^* = 4\pi d^{*2}$ where $d^*$ is the 
%value plotted in Fig.~\ref{fig:D_d} for the corresponding replica. Averaging 
Fig.~\ref{fig:proba_at_least_1} shows the sigmoid form of this probability 
as a function of $N_d$, computed by using 400 replicas. 
%over replicas gives a smooth sigmoid-shaped curve, as plotted in 
%Fig.~\ref{fig:proba_at_least_1}. 
%From this curve, 
To obtain an effective value of $d^*$ (or $N_d^*$) 
for an ensemble of many replicas, we define the
quantity $\tilde{d}^*$ (or $\tilde{N}_d^*$) by the condition that 
$\Pi^N_{\tilde{N}_d^*}=1/2$.
%We can define a representative value  $\tilde{N}_d^*$, 
%resp. $\tilde{d}^*$, as the value of $N_d$, resp. $d$, for which 
%$\Pi(\tilde{N}_d^*)=1/2$. These quantities are defined in average over replicas,
%for a given system size $N$. 
As mentioned earlier, larger systems have higher probability of having a pacemaker-region
for a fixed value of $d$ (or $N_d$) which is explicitly shown in Fig.~\ref{fig:proba_at_least_1}.
%The larger the system size, the larger the 
%probability for a fixed value of $N_d$. Stated differently, this is our
%previous observation that larger systems sustain oscillations for larger 
%values of $d$ or $D$ (Fig.~\ref{fig:A_Nosc}).

%We can express analytically the numerically defined probability of  Eq.~(\ref{eq:def:Pi}). 
If we assume that there are 
$\eta$ uncorrelated, {\it i.e.} effectively independent,
subsets of size $N_d$ in the 2D system,
%$\eta=N/N_d$ non-overlapping subsets of size $N_d$ in the 2D system, 
we have:
\begin{equation}
\Pi^N_{N_d} = 1-\left(1-P^N_{N_d}\right)^\eta.
\label{eq:P_1_pace}
\end{equation}
One possible estimate of $\eta$ is to assume that the independent subsets
are obtained by tiling the system using non-overlapping blocks of size $N_d$,
i.e., $\eta  = N/N_d$.
%The assumption that $\eta = N/N_d$, {\it i.e.} that the only independent subsets
%are strictly non-overlapping
%subsets of $N_d$ cells
%are independent,  all the subsets are assumed to be non-overlapping and therefore independent, 
However, a single pacemaker region may be split among several neighboring 
tiles, which effectively leads to an underestimation of the 
probability $\Pi^N_{N_d}$.
Another possible estimate of $\eta$ is to assume that {\it all} blocks
of size $N_d$ are independent, so that $\eta = N$.
However,  the high degree of overlap between neighbouring blocks introduces
significant correlations between them, leading to an overestimation
of $\Pi^N_{N_d}$ in
Eq.~(\ref{eq:P_1_pace}).

In general, we expect that the value of $\eta$ will be related to
$N$ and $N_d$ by $\eta = m \cdot N/ N_d$. 
The existence of a number $m$ independent of the system size can 
be qualitatively understood from the following argument.
Consider the 
covariance of the two variables, $f_d(i,j) = N_p(i,j)/N_d$ and $f_d(i',j')$, 
i.e.,
of the average number of passive cells at the two sites $(i,j)$ and $(i', j')$ 
separated by a distance $l = \sqrt{ (i - i')^2 + (j-j')^2 }$. It
has the exact expression 
$\sigma \rho(l)/ N_d$ where $\sigma=(1-\zeta) f$ and 
$\rho(l)=\exp[ -(l/2d)^2]$ 
is the correlation between the two sites. Thus, how fast the two values of $f_d$
decorrelate is simply given by $\rho(l)$. 
To estimate the number of independent
subsets containing $N_d$ points, we introduce $\kappa$, defined as the maximum
possible
correlation between two independent subsets. The correlation length $l_0$
is then defined by $\rho(l_0) = \kappa$, so that two subsets separated by 
a distance $ l > l_0$ are independent. This implies the existence of
$N/l_0^2$ independent blocks, suggesting in turn that $m = N_d/l_0^2$. 
Intuitively, one expects $l_0$ to be of the order of the width of the 
Gaussian kernel, $d$, yielding $ m \approx 4 \pi$.
We have numerically obtained the effective number $m$ by least-square
fitting of the numerical data shown in Fig.~\ref{fig:proba_at_least_1} with
the theoretical expression of Eq.~(\ref{eq:P_1_pace}). 
This gives values of $m$ lying in the range $ 9.5 \le m \le 16.5$, which
is in fact consistent with the heuristic estimate $m = 4 \pi$. 

\subsection{System size dependence of the probability of occurrence of pacemaker-like region}
\begin{figure}[tb]
\begin{center}
\includegraphics[width=0.99\linewidth]{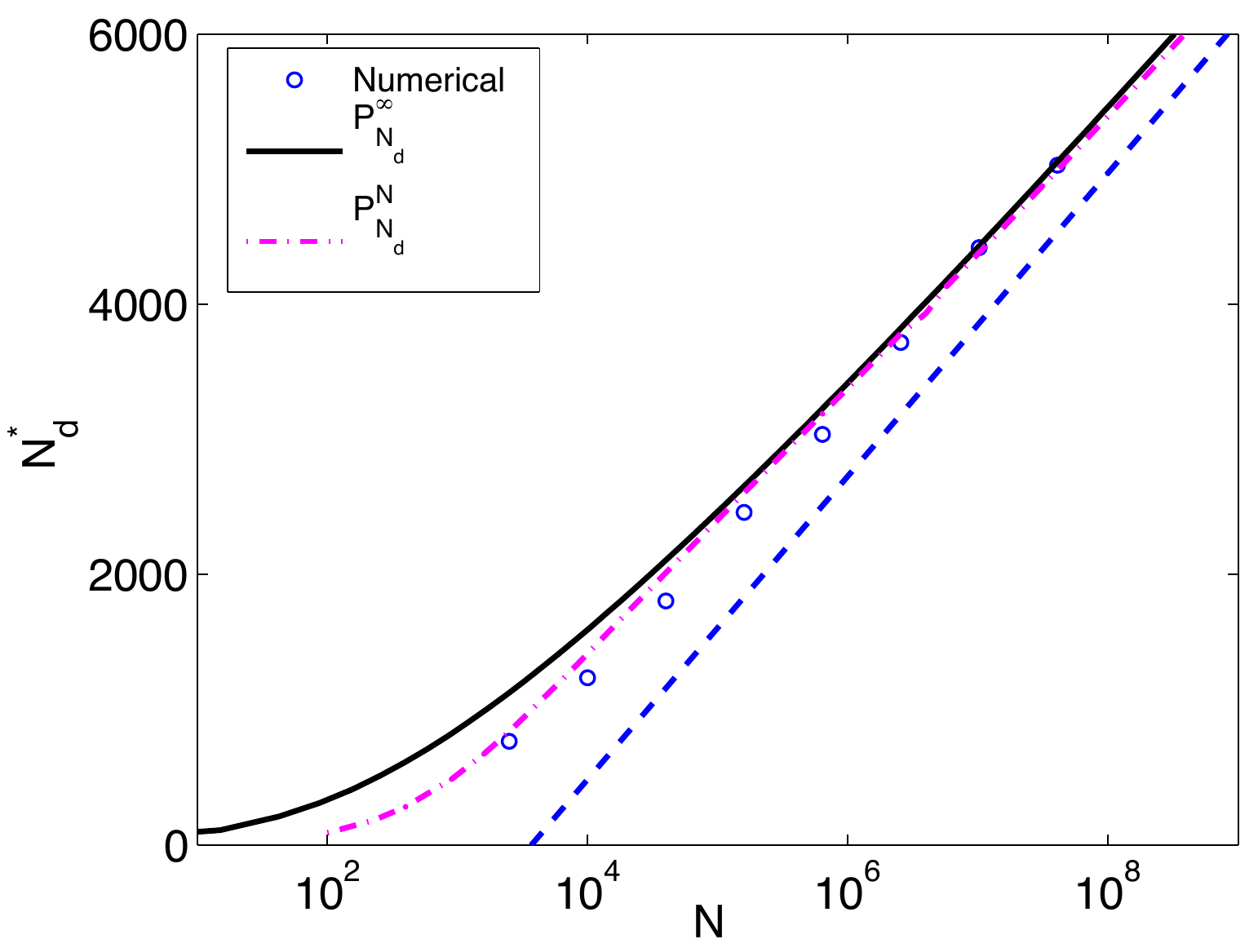}
\end{center}
\caption{
The largest coarse-graining size for which oscillations are observed in the system, $N_d^*$, defined as $\Pi^N_{N_d^*}=1/2$, as a function of the system size $N$. 
The circles represent numerical data shown in Fig.~\ref{fig:proba_at_least_1}.
The dashed straight line is a guide to the eye indicating the slope 
$1/(\lambda f_c^l)$. The solid line 
and dashed-dotted line represent the solutions of Eq.~(\ref{eq:P12}) using the 
expressions given by Eq.~(\ref{eq:PNd_finite}) and Eq.~(\ref{eq:PNd_infinite}) respectively.}
\label{fig:P_1}
\end{figure}

We now obtain the cutoff value $N_d^*$ for the coarse-graining size below 
which oscillations are present in the system by solving 
\begin{equation}
\Pi^N_{N_d^*} = 1-\left(1-P^N_{N_d^*} \right)^{m\frac{N}{N_d^*}}=1/2.
\label{eq:P12}
\end{equation}
To solve this equation, we assume that the system is large enough, so that 
$P^N_{N_d^*}$ can be replaced by $P^\infty_{N_d^*}$,
Eq.~(\ref{eq:P_Nd_simple}).
This implies that the only system-size dependence of $\Pi^N_{N_d}$ 
is from the exponent $mN/N_d$. We thus obtain a linear relation between 
$\log(N)$ and $N_d^*$ (Fig.~\ref{fig:P_1}):
\begin{equation}
\mu^2 f_c^l N_d^* = 2 \log N + 2 \log \frac{\mu^2 f_c^l m}{\log(2)\sqrt{2\pi}},
\label{eq:correction_sublog}
\end{equation}
which effectively defines $N_d^*$.
For large systems, the behavior of $N_d^*$ is well described by the 
$\log(N)$ term
in Eq.~(\ref{eq:correction_sublog}) shown by a straight line in 
Fig.~\ref{fig:P_1}. 
It is worth noting that the $\log(N)$ scaling appears very naturally in 
the general 
context of extreme value statistics~\cite{Maj_JSP_2005}. 
The constant term in the R.H.S. of 
Eq.~(\ref{eq:correction_sublog}) is responsible for the deviations from
$\log(N)$ scaling seen for $N \lesssim 10^4$.

\section{Scaling in the transition to activity in spatially extended systems}
\begin{figure*}[tb]
\begin{center}
\includegraphics[width=0.98\linewidth]{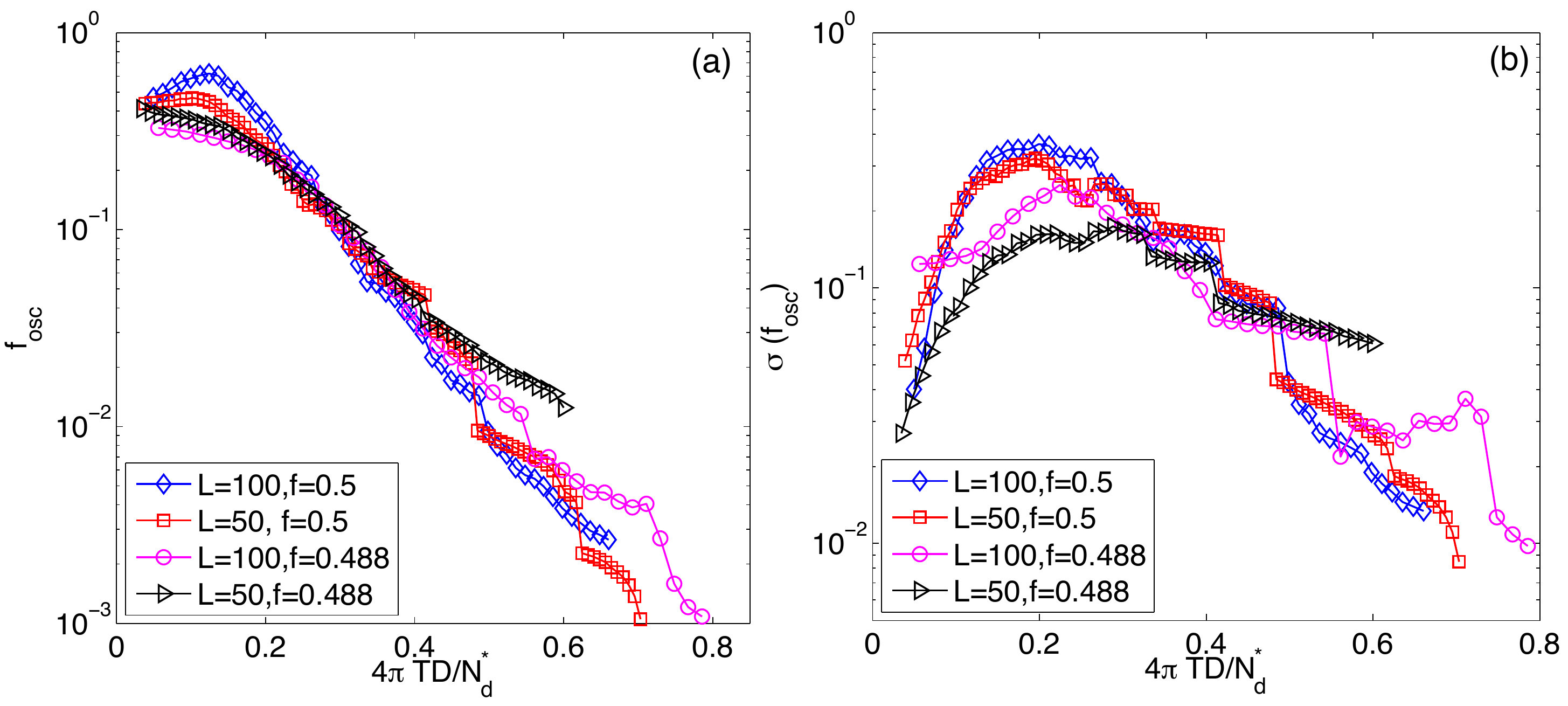}
\end{center}
\caption{
(a) Average value and (b) standard deviation of the fraction of oscillating 
cells in a 2-D system, shown as a function of the reduced variable 
$4 \pi D T /N_d^*$, where $N_d^*$
% using the same data
%as in Fig.~\ref{fig:A_Nosc}, plotted as
%a function of the reduced variable $N_d/N_d^*$, where $N_d^*$
is defined in Eq.~(\ref{eq:correction_sublog}).
The symbols indicated in the legend correspond to different 
values of $L$ and $f$, {\it i.e.}, $N$ and $\mu$.
The different
curves collapse to the same form (within statistical accuracy) when
$N_d$ is large, the regime for which the 
analysis described in the text is valid.
%Open symbols : $f=0.488$, closed symbols : $f=0.5$, black : $N_x=100$, brown : $N_x=50$. 
%Left : first transition at $f_c^l$; right : second transition at $f_c^u$.
%{\color{red} to-do : same units in x / same ylimits.}
}
\label{fig:renormalized}
\end{figure*}

%\subsection{Influence of the mean passive cell density}
%In the phase diagrams (Fig.\ref{fig:Phases}), we focused on the two coupling parameters $C_r$ and $D$. 
%In the mean-field limit, the average passive cell density $f$ plays the same role as $C_r$, whereas in $2$-d system $f$ gets %averaged by diffusion. 
 The analysis presented in the previous section
%, in particular 
%Eq.~\ref{eq:correction_sublog} and Fig.~\ref{fig:P_1}, 
allows us to identify a characteristic coarse-graining size, suggesting
% which determines the existence of
%pacemakers in a system of size $N$. 
%Our analysis suggests 
that the number of pacemaker-like regions
in a given system is determined by the ratio $N_d/N_d^*$,
where $N_d^*$ is given by Eq.~(\ref{eq:correction_sublog}). 
In addition, Fig.~\ref{fig:D_d} shows that 
diffusive coupling $D$ results in coarse-graining of the passive cell density
over a region of size 
%where $N_d $ is proportional to $D$:
$N_d \approx 4 \pi D T$, with $T$ being the typical oscillation period 
(subsection~\ref{subsec:emergence}). 
%Thus the number of
%pacemaker-like regions in a given system is a function of
%$4 \pi D T/N_d^*$.
%where $N_d^*$ is given by Eq.~(\ref{eq:correction_sublog}). 
Based on this, we expect that the fraction of oscillating cells
will depend on the reduced variable  
$4 \pi D T/N_d^*$.

Fig.~\ref{fig:renormalized} shows the average (a) and the 
standard deviation (b) of the fraction of oscillating cells, $f_{osc}$,
averaged over many ($\sim 100$) replicas as a function of 
$4 \pi D T/N_d^*$, for several values of $f$ (or equivalently, of $\mu$) 
and $L$. The curves for the mean value of $f_{osc}$ 
seem to collapse to a common form at large $N_d$ where our analysis applies,
Eq.~(\ref{eq:P_Nd_simple}), is not valid when $N_d$ is small;
see in particular the discussion at the end of subsection~\ref{subsec:5.2}),
supporting the conclusion that the properties of the system 
indeed vary with the scaling parameter $D/N_d^*$. The deviations seen in 
the different curves can be explained as an outcome of the large standard
deviation in $f_{osc}$ as seen from Fig.~\ref{fig:renormalized}~(b).
Remarkably, apart from the curves corresponding to the average values, we also observe relatively good collapse in the curves corresponding to the standard deviation of $f_{osc}$, at least when $D /N_d^*$ is not too
small. This suggests that our analysis may apply not only to 
the averaged value, but also to the second moment of the distribution of 
the fraction of oscillating cells, $f_{osc}$.

As mentioned in the previous section, in the limit of very large system size, 
$N_d^*$
is a function of the reduced variable $\mu^2 D/\log(N)$. 
This scaling implies that for a fixed value of the coupling between excitable 
cells, as system size is increased and/or the mean value $f$ of the passive 
cell distributions comes closer to the critical value $f_c^l$, there is higher probability of observing sustained oscillatory activity in the system.
Alternatively, if $f$ is kept fixed at a value smaller than $f_c^l$, then 
the larger
the system size, the larger the range of values of $D$ over which oscillations
can be observed. 

\section{Conclusion}

In this paper we explore the role of heterogeneity in the spatial 
distribution of 
passive cells, which introduces quenched disorder in a system
of coupled excitable and passive cells, on the transition between quiescent 
state and oscillatory activity. 
%in a system of coupled excitable cells,
%each of which is connected to a variable number of passive cells.
%(rajeev)
%This has been achieved by several steps. The first step has been to study
%the simplest set of an excitable cell, couple to several passive cells,
%without any geometry ($0$-d system).
We begin our analysis with the mean-field limit of the spatially
extended system, 
%corresponding to extremely coupled excitable cells, 
which is equivalent to a single excitable cell 
coupled to a number of passive cells. We observe here that
the transition from quiescent to oscillatory activity is subcritical, 
i.e., oscillations appear at onset with a finite amplitude. 
We next investigate the system at finite values of the coupling between 
excitable
cells and characterize the coarse-graining effect of diffusion. We
show that the dynamical 
%follow} this up with a study of spatially extended systems, and
%study the coarse-graining effect of diffusion 
%by showing that the dynamical 
behavior of the system is related 
to the local passive cell density obtained using a suitable 
coarse-graining length scale.  
We observe large variability in the dynamical behavior of different
replicas, i.e., different realizations of 
the spatial distribution of passive cells. The strong effect
of this quenched disorder on the dynamics is reminiscent of 
glassy systems.
%Increasing this length scale eventually
%results in the system exhibiting the mean-field behavior.
To characterize the properties averaged over many
realizations of the disorder, we assume
that the effect of diffusion of strength 
$D$ is to coarse-grain the local passive cell distribution
over $N_d$ sites. 
We indeed empirically establish that 
$D \propto N_d $.
% with the constant of proportionality being determined
%by a combination of {\bf heuristic considerations ?} and fitting. 
From this perspective, we analyze the transition between quiescent state
and sustained oscillations.
We obtain
a scaling relation that describes the transition as a function of (i) system
size, (ii) coupling between excitable cells and (iii) average passive cell 
density.
%We quantified the influence of the system size as facilitating the 
% existence of oscillations and showed that this effect is logarithmic in $N$.
One of the important implications of this relation is that the occurrence of
oscillatory behavior depends on
the logarithm of the system size $N$ so that increasing $N$ enhances
the probability of observing oscillations.
%We note that our numerical study has been carried out for one particular value
%of $C_r$; in this respect it would be interesting to study the dependence of 
%$T$ as a function $C_r$ (and/or $K$).  (do we need to say this? rajeev)
% (no new para, rajeev)
%As a consequence of these properties, governing the probability to observe 
%pacemaker-like regions in the system, we have established the 
%dependence of the transition towards a regime of activity as a function of 
%the size $N$ of the system. 
%The $\log(N)$ scaling of the control parameter is observed both for
%very large systems (Poisson distributions) and intermediate systems
%sizes (binomial distribution). 
%Apart from increasing system size,

As mentioned in the introduction, the model system that we have
analyzed has been motivated by the biological phenomenon of onset of
coherent oscillations in the pregnant uterus close to term.
One of the implications of our work is that larger organs may show 
greater variability in their dynamical behavior for a given set of
parameters describing the state of the system.
In particular, they may be more likely to exhibit oscillations 
even prior to the transition point as a result of spatial fluctuations, 
potentially implying that mammals having bigger uteri will be at 
higher risk of having pre-term rhythmic activity. 
As it is not yet well-understood why in some cases periodic dynamical
behavior is initiated in uterine tissue significantly earlier than
normal, our study of the role of disorder in creating an effective
pacemaker-like region giving rise to rhythmic activity in such systems
may be of potential significance for possible clinical applications. 
Our work also connects the dynamical phenomena seen in
such biological systems with the study of the role of disorder in
phase transitions occurring in several physical systems, including
spin glasses.

\medskip

\noindent
%{\bf Acknowledgement} 
We are grateful to E. Bertin, P. Leboeuf and S. Majumdar for insightful
discussions. We thank the HPC facility (IMSc) and the PSMN (ENS Lyon) for 
providing computer resources. This research was supported in part by
IFCPAR (Project No. 3404-4).

%\bibliography{FHN_bib}

%\section*{References}

\end{document}